\documentclass[11pt,a4paper,dvips]{article}

\usepackage{amstex}
\usepackage{latexsym}
\usepackage{graphicx}
\usepackage{amssymb}

\def\etal{et al.}
\def\delv{\Delta v}
\def\Lya{Ly$\alpha$ }
\def\lya{Ly$\alpha$ }
\def\ndmp{17}
\def\nrun{8,500}
\def\kms{km~s$^{-1}$ }
\def\cm2{\, \rm cm^{-2}}
\def\N#1{{N({\rm #1})}}
\def\f#1{{f_{\rm #1}}}
\def\rAA{{\rm \, \AA}}
\def\sci#1{{\rm \; \times \; 10^{#1}}}
\def\ltk{\left [ \,}
\def\ltp{\left ( \,}

\def\rtk{\, \right  ] }
\def\rtp{\, \right  ) }

\def\ohf{{1 \over 2}}

\def\rhf{{3 \over 2}}

\def\perd{\;\;\; .}
\def\cmma{\;\;\; ,}

\def\Nperp{N_{\perp} (0)}
\def\intl{\int\limits}
\def\und#1{{\rm \underline{#1}}}
\def\apj{ApJ}
\def\apjsupp{ApJS}
\def\mnras{MNRAS}
\def\nat{Nature}
\def\araa{ARAA}
\def\aj{AJ}
\def\aap{A $\&$ A}

\title{ON THE KINEMATICS OF THE  DAMPED \lya\ PROTOGALAXIES}

\author{ JASON X. PROCHASKA$^1$
\& ARTHUR M. WOLFE$^1$
\\
Department of Physics, and Center for Astrophysics and Space Sciences; \\
University of California, San Diego; \\
C--0424; La Jolla; CA 92093\\}

\begin{document}
\maketitle

\begin{abstract} 

\footnotetext[1]{Visiting Astronomer, W.M. Keck Telescope.
The Keck Observatory is a joint facility of the University
of California and the California Institute of Technology.} 

We present the first results of an ongoing program
to investigate the kinematic properties of high redshift
damped \lya systems.  Because damped \lya systems are widely believed
to be the progenitors of current massive galaxies, 
an analysis of their kinematics  allows
a direct test of galaxy formation scenarios. Specifically,
the kinematic history of protogalactic gas is a sensitive
discriminator among competing theories of galaxy formation.

We use the HIRES echelle spectrograph on the Keck
10 m telescope to obtain accurate, high-resolution spectra of 
17 damped {\lya} systems.  We focus on 
unsaturated, low-ion transitions such as
Si II 1808, since these accurately trace the velocity fields
of the neutral gas dominating the baryonic content of the
damped systems. The velocity profiles: (1) comprise multiple
narrow components; (2) are asymmetric in that the component
with strongest absorption tends to lie at one edge of the profile;
and (3) exhibit a nearly uniform distribution of velocity widths 
between 20 and 200 {\kms}.

In order to explain these characteristics, 
we consider several physical models proposed
to explain the damped \lya phenomenon,
including rapidly rotating ``cold'' disks, slowly rotating ``hot'' disks,
massive isothermal halos, and a hydrodynamic spherical accretion model.
Using standard Monte Carlo techniques, we run sightlines through
these model systems to derive simulated low-ion profiles.
We develop four test statistics
that focus on the symmetry and velocity widths of the profiles
to distinguish among the models.
Comparing the distributions of test statistics from the simulated profiles
with those calculated from the observed profiles, we determine that the
models in which the damped {\lya} gas is distributed 
in galactic halos  and in spherically infalling gas,
are ruled out at more than 99.9$\%$ confidence. A model
in which dwarf galaxies are simulated by slowly rotating ``hot'' disks
is ruled out at 97$\%$ confidence.
More important, we demonstrate that the Cold Dark Matter Model,
as developed by Kauffmann (1996), is inconsistent with the damped
\lya data at more than the 99.9$\%$ confidence level. This is because
the CDM Model predicts the interception cross-section of damped
{\lya} systems to be dominated by systems with rotation speeds
too slow to be compatible with the data. This is an important result,
because slow rotation speeds are generic traits of protogalaxies in
most hierarchical cosmologies.

We find that models with disks that rotate rapidly and are
thick are the only {\it tested}
models consistent
with the data at high confidence levels. 
A Relative Likelihood Ratio Test
indicates disks 
with rotation speeds, $v_{rot} < 180$ {\kms}, and 
scale heights, $h$ $<$ less than 0.1 times the radial scale
length $R_{d}$, are ruled out at the 99$\%$ confidence
level. 
The most likely 
values of these parameters are $v_{rot}$ = 225 {\kms} and
$h$ = 0.3$R_{d}$. We also find that these disks must
be ``cold'', since models in which $\sigma_{cc}$/$v_{rot}$ $>$ 0.1
are ruled out with 99$\%$ confidence, where $\sigma_{cc}$ is
the velocity dispersion of the gas.
We describe an independent test of the ``cold'' disk hypothesis. The
test makes use
of the redshift of emission lines sometimes detected in damped 
{\lya} systems, as well as the
absorption profiles. The test potentially distinguishes among
damped systems that are: (1) large rotating disks
detected in absorption and emission, in which case a
systematic relation exists between emission redshift and absorption velocity
profile, and (2) emitting galaxies surrounded by satellite galaxies 
detected in damped {\lya} absorption, in which case the relation
between emission and absorption redshifts is random. 

Finally we emphasize a dilemma stimulated by our findings. Specifically,
while the kinematics of the damped {\lya} systems strongly favor a ``cold''
disk-like
configuration, the low metallicities and type II Sn abundance patterns
of damped {\lya} systems
argue for a ``hot'' halo-like configuration.
We speculate on how this dilemma might be resolved.  \end{abstract}

keywords cosmology---galaxies: evolution---galaxies: 
quasars---absorption lines 

\clearpage

\section{INTRODUCTION}

Velocity fields within protogalaxies carry
important clues about the process of galaxy formation.
The kinematic history
of the protogalactic gas is of particular importance as
it is a sensitive discriminator between competing theories
of galaxy formation. In standard
hierarchical models, low-mass subunits of dark matter and gas continuously
merge to form the high-mass galaxies seen today \cite{wht91}. In 
this scenario, the rotation speeds of galaxy disks at 
high redshifts ($z \gtrsim 2.5$) are low compared to the rotation 
speeds of current massive spirals \cite{kau96}. By contrast, 
higher rotation speeds are predicted at large
redshifts by
models in which disks form from the coherent collapse of gas in
massive dark-matter halos \cite{egg62,fal80}. 

The purpose of this paper is to test these ideas {\em directly}
with data that accurately measure the motions of protogalactic gas.
Specifically, we present an analysis of high-resolution spectra
of metal absorption lines from a sample of damped {\lya} systems. The
latter are high-redshift layers of neutral gas 
widely thought to be the progenitors of current 
galaxies \cite{wol95b,pbl93,wht91,fug96}.
We use the HIRES Echelle spectrograph \cite{vgt92} 
on the Keck 10 m telescope to detect the gas
in absorption against the light of more  distant QSOs.
The velocity profiles of the damped systems are notable for 
the following properties: (1) they comprise multiple narrow
components; (2) they
are asymmetric in that the component with strongest absorption tends
to be at one edge of the profile; and (3) the velocity intervals
over which absorption occurs, $\Delta v$, are uniformly distributed
between $\approx$ 20 and 200 {\kms}.
By constructing statistical measures characterizing the 
velocity structure of the absorption profiles, we 
demonstrate that the asymmetry rules out models dominated by
random motions or by spherically symmetric radial motions. Rather, we show 
the asymmetry can be caused by the rotation of ``cold'' gaseous
disks with significant vertical scale heights.  Furthermore,
the distribution of velocity 
intervals implies high rotation speeds, $v_{rot}$ $\approx$ 225 {\kms}.

Two types of models have been proposed to explain
the damped {\lya} systems.  The hierarchical
schemes assume the damped systems to be slowly rotating disks
embedded in low-mass protogalaxies
\cite{kly95,mom94,ma94,kau96}.  Other models,
that do not assume a particular cosmological context, propose the
damped systems to be rapidly rotating disks in high-mass
protogalaxies \cite{sch90}, 
protospheroidal gas \cite{lu96b},
dwarf galaxies,
\cite{yor86}, or gas falling into
protogalaxies \cite{arn72}.
We compare all the model predictions with
the data using Monte Carlo techniques. 
The models are constructed by inserting gas
into discrete velocity components according to 
algorithms described below. We randomly 
select the locations and velocities of components along the line
of sight from parent distributions appropriate for the scenario
being tested. Synthetic spectra are generated by computing
the absorption profiles for optically
thin transitions arising in low ions such
as Si$^{+}$, because the latter accurately traces the velocity structure
of the neutral gas.
We generate an ensemble of velocity profiles by assuming the 
model protogalaxies
are randomly oriented in projection on the sky. We then
compare the computed and observed distributions of statistical measures
that are sensitive to the degree of asymmetry and width
of the absorption profiles.  As demonstrated below,
the asymmetry in the velocity profiles rules out ``hot'' models in
which $\sigma_{cc}$ $\gtrsim$
$v_{rot}$ at high confidence levels, where
$\sigma_{cc}$ is the
component-component velocity dispersion. This is why models
with significant random motions are unlikely. 
We then focus on ``cold'' rotating disks
in which $\sigma_{cc}$ $<<$ $v_{rot}$
and compare the model distribution of 
$\Delta v$ with the observed distribution. 
This particular test is crucial for
evaluating model disks with different rotation speeds and is
the test that rules out models with low $v_{rot}$.
In fact, {\em the principal conclusions of this
paper is that low rotation speeds, characterizing the bulk of
protogalactic disks, in most hierarchical cosmologies, are
ruled out at high confidence levels.  Rather, protogalactic
disks appear to rotate rapidly.}

The paper is organized as follows. In $\S$ 2 we discuss the
criteria used to select metal-line absorption profiles suitable
for model testing. 
We select a sample of low-ion profiles
from {\ndmp} damped {\lya} systems.  
Optical depth spectra are constructed that will be used
for tests discussed in the following sections.
In $\S$ 3 we introduce the rotating disk model and discuss the
Monte Carlo methods. 
Here we determine input parameters used to generate the
simulated spectra. We discuss extrinsic parameters such as
the number of velocity
components, the velocity and position of each component, $\sigma_{cc}$, etc. 
We also discuss the  selection of intrinsic parameters such
as the internal velocity dispersion, the column density
per component, etc. Together with various model assumptions
the parameters are used to create
simulated spectra with the noise and spectral resolution of the
data.    The isothermal halo, spherical infall and random models
are discussed in $\S$ 4.  The statistical
tests are described in $\S$ 5 where attention is given to
statistical measures devised to discriminate among the
models.  In $\S$ 6 we present the results of these tests, in particular
demonstrating the failure of all of the models except the ``cold'',
rapidly
rotating disk model.  We explore the allowed parameter space
for the latter model in $\S$ 7 and perform
the Likelihood Ratio Test to determine the most likely set of 
values for $v_{rot}$ and the thickness of the disks.  An
independent test of the rotating disk hypothesis is
presented in $\S$ 8. Concluding remarks are  
made in $\S$ 9.

\section{PROFILE SELECTION}

This section discusses the  sample of
spectra used for model testing. We construct
apparent optical-depth spectra
and apply various criteria to them in order
to select transitions suitable for testing.

\vspace{0.3in}

\subsection{Data}

     The data analyzed in this paper were acquired with the
High Resolution Spectrograph (HIRES) located at the Nasmyth
focus of the 10m W. M. Keck Telescope \cite{vgt92}.  
Table~\ref{data} lists the
coordinate name of the background QSO, 
observation date, exposure time, emission redshift, resolution
and signal to noise ratio (SNR) of the damped \Lya systems
comprising two data sets.  The first data set (W) comprises
observations performed by one of us (AMW) as part of a long term project
to measure the metallicity and kinematics of damped \lya systems.
With the exception of Q1331+170, the C5 decker,
that has a FWHM resolution of $\approx$ 8 \kms,
was used for all
of the observations.
For Q1331, we used the C1 decker that gives a higher FWHM resolution of 
$\approx 6$ \kms.  The latter resolution and setup was used by W. L. W. Sargent
who generously provided the second data set (S).

\begin{table*}
\caption{QSO AND OBSERVATIONAL DATA} \label{data}
\begin{center}
\begin{tabular}[h]{lcccccl}
\hline
\hline
QSO & Date
& Exposure
& $z_{em}$
& Resolution & SNR & Data \\
& & Time (s) & & (\kms)& & set \\
\hline
Q0100+1300 & S94 & 11700 & 2.69 & 7.5 & 40 &W \cr
Q0201+3634 & F94 & 34580 & 2.49 & 7.5 & 35  &W\cr
Q0458$-$0203 & F95 & 28800 & 2.29 & 7.5 & 15  &W\cr
Q1215+3322 & S94 & 14040 & 2.61 & 7.5 & 20  &W\cr
Q1331+1704 & S94 & 36000 & 2.08 & 6.6 & 80  &W\cr
Q2206$-$1958A,B & F94 & 25900 & 2.56 & 7.5 & 40  &W\cr
Q2231$-$0015 & F95 & 14400 & 3.02 & 7.5 & 30  &W\cr
Q0216+0803 & F94,F95 & 18600 & 2.992 & 6.6 & 24&S$^{a}$ \cr
Q0449$-$1325 & F94 & 16000  & 3.097 & 6.6  & 40 &S\cr
Q0528$-$2505$^{b}$ & F94,S95 & 24000 & 2.779 & 6.6 & 25 &S\cr
Q1202$-$0725 & S95 & 33000 & 4.7 & 6.6 & 30 &S\cr
Q1425+6039 & S95 & 37200 & 3.173 & 6.6 & 100 &S\cr
Q1946+7658A,B & S94 & 8700 & 2.994 & 6.6 & 55 & S\cr
Q2212$-$1626 & F93,F95 & 33000 & 3.992 & 6.6 & 75&S \cr
Q2237$-$0608 & F94, F95 & 56000 & 4.559 & 6.6 & 22&S \cr
\end{tabular}
\end{center}

$^{a}$  Data kindly provided by W. L. W. Sargent and collaborators 

$^{b}$  The $z=2.811$ damped system toward Q0528$-$2505
was omitted because its redshift exceeds that of the background QSO to which
it may be associated. 

\end{table*}

     All of the data were reduced with a software
package kindly provided by T. Barlow.  
Briefly, the routine optimally extracts the data from
its 2-D CCD format into 1-D spectra by using a bright standard 
star image as a trace.  The package properly flat fields and 
wavelength calibrates the data with the quartz and ThAr calibration
frames while removing cosmic rays.  
Finally, the data is continuum fitted  with the IRAF
package {\it continuum}.  The reduction process 
is both accurate and relatively
free of systematic errors.  Thus the data sample is essentially homogeneous
with the only variations arising from differences in SNR and resolution.
The sample is also unbiased, because the objects were selected without
{\it a priori} knowledge of the kinematic structure of the absorbing gas.
Rather, the objects were selected for: (1) large H I column 
density\footnote{The exceptions to this rule are the $z = 1.267$
system toward Q0449$-$135 and the 
$z = 1.738$ system toward
1946$+$770 . The low-ion column densities in both systems are
large enough to indicate
H I column densities, that have not
been measured directly from {\lya},  
exceeding the threshold $\N{HI} = 2\sci{20} \cm2$.}
$N$(H I) $\ge$ 2$\times$10$^{20}$ cm$^{-2}$, and
(2) brightness of the background QSO, $V$ $\le$ 19.  

\begin{figure}
\includegraphics[height=8.0in, width=6.5in]{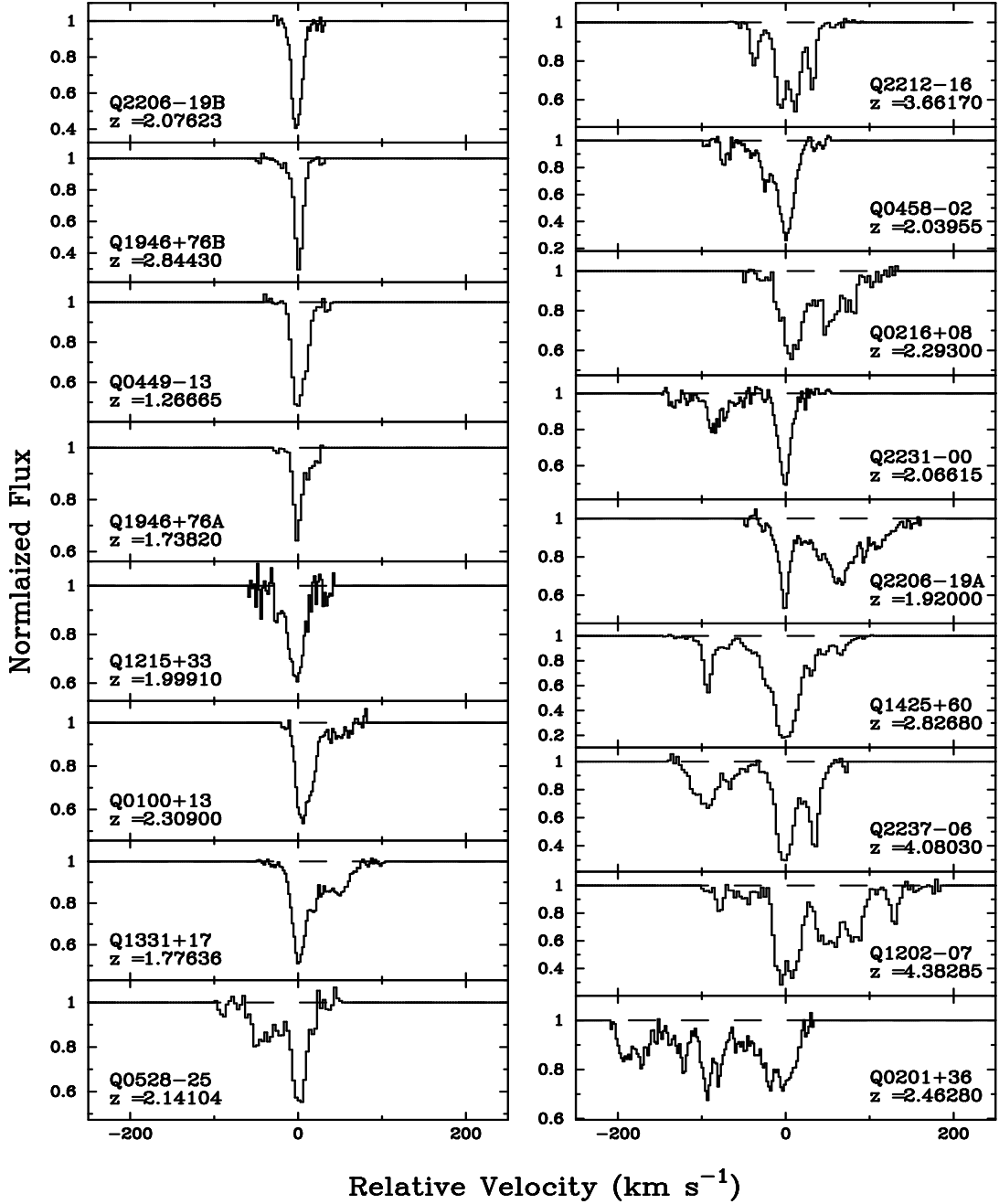}
\caption{Velocity profiles of low-ion transitions from
17 damped \lya systems comprising our empirical sample.  
For each profile, $v = 0$ \kms corresponds to the redshift
labeled in the plot.}
\label{sptra}
\end{figure}

\begin{figure}
\includegraphics[height=7.5in, width=6.5in]{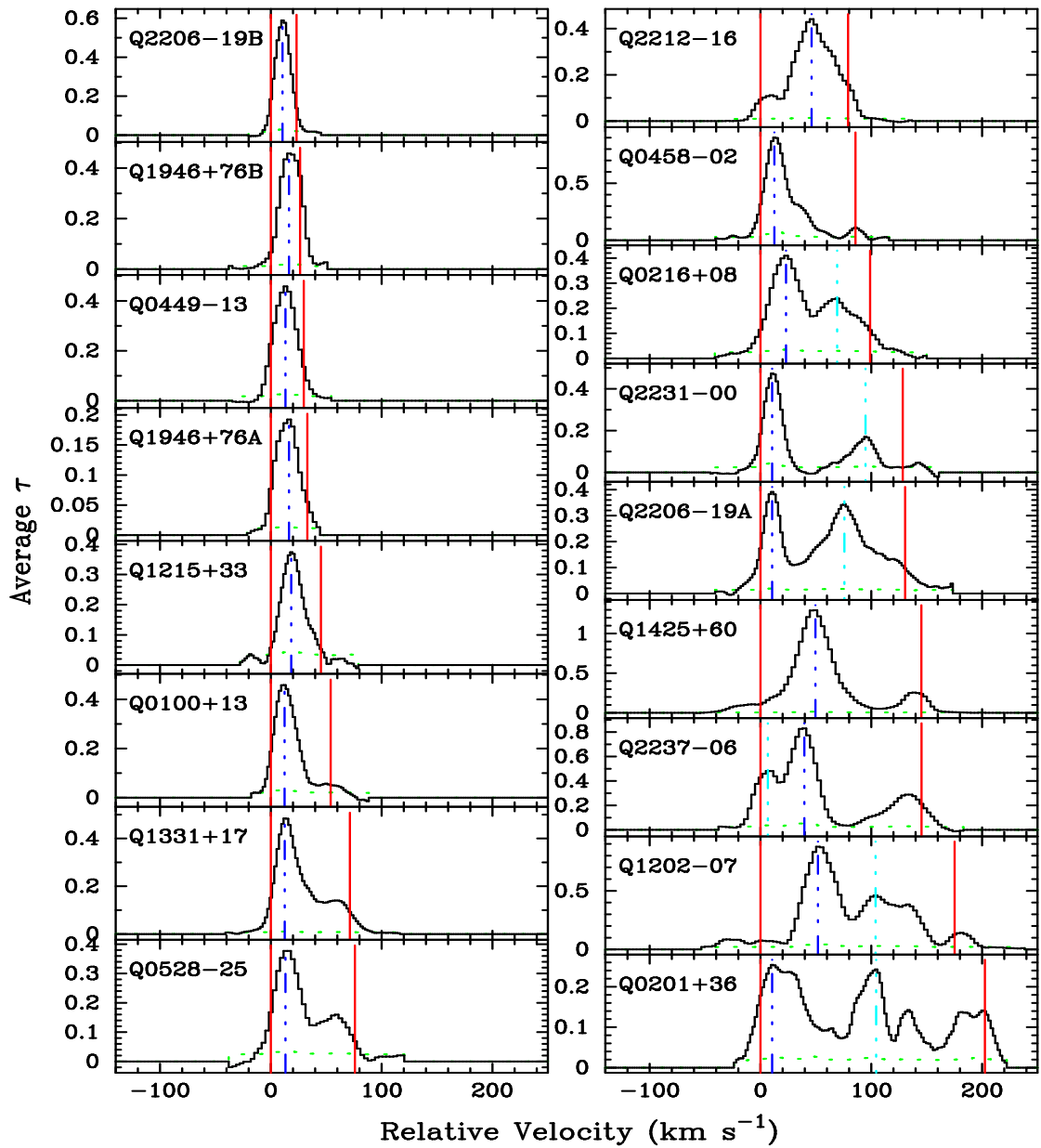}
\caption{Average optical depth profiles (binned over 9 pixels)
for the same 17 transitions presented in Figure 1.
The solid vertical lines designate the velocity interval, $\delv$,
while the dash-dot lines indicate peaks and the dotted line is the
$2 \sigma(\bar\tau)$ array.  
The profiles have been shifted in velocity space such
that the left edge of the velocity interval coincides with $v=0$ and
reflected if necessary
so that the strongest peak always lies on the left edge of
the interval.}
\label{obstau}
\end{figure}

\subsection{Profile Selection Criteria}
\label{trncrit}

In order to form an ensemble of velocity profiles  we 
select the profile of a single transition from  each damped {\lya} system.
We first require that the line profile be free from blending with other
absorption line profiles.  This condition is met by intracomparing
a set of three or more transitions from each damped {\lya} system.
We next construct an apparent optical-depth spectrum, 

\begin{equation}
\tau(v) \equiv \ln [I_{c}/I(v)] \cmma
\end{equation}

\noindent where $I(v)$ and
$I_{c}$ are the transmitted and incident intensities at velocity $v$.
To focus on spectral features rather than individual pixels, 
we create a binned optical depth array by averaging 
$\tau(v)$ over a running window nine pixels wide:  

\begin{equation}
{\bar \tau(v_{i})} = {1 \over 9} {\sum_{j=i-4}^{i+4}{\tau(v_{j})}} \perd
\end{equation}

\noindent 
In the last equation $\tau(v_{j})$ is the apparent optical
depth at the $j^{th}$ velocity pixel and 
the sum is over the four  nearest neighbors of pixel $i$ inclusive.  

In order to assure the detection of weaker features, which is crucial
for establishing asymmetries in the line profiles,
we demand that
$\bar\tau(v_{pk})$, the $\bar\tau(v)$ of
the strongest feature (i.e. the feature with peak optical depth),
exceed 20$\sigma (\bar\tau)$ significance.
Specifically, with this criterion we can detect
features  with $\bar\tau(v)$ $\ge$ 0.25$\times$$\bar\tau(v_{pk})$
at 5$\sigma$ significance.
For spectra with SNR $\approx$ 30 this criterion implies
the strongest feature must satisfy $I(v_{pk})/I_{c}$ $\leq$ 0.6.
Because saturated
lines are strongly affected by opacity that can mask asymmetries
and component structures detectable in optically thin
transitions, we set an upper limit on $\bar\tau(v)$ by
demanding $I(v_{pk})/I_{c}$ $\ge$ 0.1. 
Therefore, we require 0.1 $\le$  $I(v_{pk})/I_{c}$ $\le$ 0.6
and note the upper limit will vary with SNR.  Given a single
transition satisfying this criterion, the results are
independent of the ionic species or the
metallicity of the damped systems, provided the
metals trace the HI gas (i.e., the system is well mixed). 

Our analysis focuses on low-ion transitions arising
in gas giving rise to damped {\lya} absorption lines. Ions 
such as Fe$^{+}$, Si$^{+}$, and Ni$^{+}$ are accurate tracers of
neutral hydrogen when $N$(H I) $\ge$ 2$\times$10$^{20}$ cm$^{-2}$,
because the
optical depth at the Lyman limit, $\tau_{LL}$ $>$ 10$^{3}$. Moreover, 
the large ratios of Al$^{+}$/Al$^{++}$ and Si$^+$/Si$^{3+}$
demonstrate the 
gas containing low ions is mainly neutral
(Prochaska \& Wolfe 1996, 1997). The low-ion transitions
selected for analysis are given in column 4 in Table~\ref{trans}. The
table also gives the QSO name in column 1, the absorption redshift
in column 2, the H I column density in column 3, the ratio 
$\bar\tau (v_{pk}) / \sigma (\bar \tau)$ in column 5, 
and the reference for each of the {\ndmp} systems in column 6. 
The corresponding intensity profiles are presented in Figure~\ref{sptra}.
The binned optical-depth spectra constructed from these profiles
are shown in Figure~\ref{obstau}.
If necessary, we reflect the profiles to place
the peak component at lower velocities.

\begin{table*}
\caption{DAMPED LYMAN ALPHA STATISTICAL SAMPLE} \label{trans}
\begin{center}
\begin{tabular}{lccccc}
\hline
\hline
QSO & $z_{abs}$
& $\log [\N{HI}]$ & Transition & $\tau_{pk}$/$\sigma_{\tau}$ & Ref. \\
& & cm$^{-2}$& & & \\
\hline
Q0100+1300 & 2.309 & 21.4 & Ni II 1741 & 29 & 1 \cr
Q0201+3634 & 2.462 & 20.4 & Si II 1808 & 21 & 2 \cr
Q0458$-$0203 & 2.040 & 21.7 & Cr II 2056 & 26 & $-$ \cr
Q1215+3322 & 1.999 & 21.0 & Si II 1808 & 20 & $-$ \cr
Q1331+1704 & 1.776 & 21.2 & Si II 1808 & 92 & $-$ \cr
Q2206$-$1958A & 1.920 & 20.7 & Ni II 1741 & 41 & 3 \cr
Q2206$-$1958B & 2.076 & 20.4 & Al II 1670 & 41 & 3 \cr
Q2231$-$0015 & 2.066 & 20.6 & Si II 1808 & 22 & 4 \cr
\cr
Q0216+0803 & 2.293 & 20.5 & Si II 1808 & 23 & 4 \cr
Q0449$-$1325 & 1.267 & N/A & Fe II 2249 & 34 & 4 \cr
Q0528$-$2505 & 2.141 & 21.0 & Si II 1808 & 22 & 4 \cr
Q1202$-$0725 & 4.383 & 20.6 & Si II 1304  & 40 & 5 \cr
Q1425+6039 & 2.827 & 20.3 & Fe II 1608 & 90 & 4 \cr
Q1946+7658A & 2.843 & 20.3 & Si II 1304 & 50 & 4,6  \cr
Q1946+7658B & 1.738 & N/A  & Si II 1808 & 29 & 4,6 \cr
Q2212$-$1626 & 3.662 & 20.2 & Si II 1304 & 65 & 4 \cr
Q2237$-$0608 & 4.080 & 20.5 & Al II 1670 & 32 & 4 \cr
\end{tabular}

$^{1}$ \cite{wol94}

$^{2}$ \cite{pro96}

$^{3}$ \cite{pro97}

$^{4}$ \cite{lu96b}

$^{5}$ \cite{lu96a}

$^{6}$ \cite{lu95a}

\end{center}
\end{table*}

The spectra in Figures~\ref{sptra} and \ref{obstau}
are notable for several reasons. 
First, the unbinned spectra in Figure~\ref{sptra} show the profiles
comprise multiple, narrow velocity components.
Second, the velocity intervals over which absorption occurs are uniformly
distributed between $\approx$ 20 and 200 {\kms}. 
Third, the velocity profiles in Figure~\ref{sptra}
are asymmetric in that the location of the strongest component tends to
lie at one edge of the velocity profile in 12 out of {\ndmp} cases.
Fourth, the binned spectra in Figure~\ref{obstau}
show evidence for a monotonic sequence of velocity components in which
optical depth decreases with increasing velocity.
To be successful, a kinematic model must quantitatively
explain these characteristics in a natural way.  

\section{ROTATING DISK MODELS}
\label{mdls}

Rotating disks have figured prominently in 
damped {\lya} research.
The original survey for damped {\lya} lines
was initiated in order to find rotating 
galactic disks at high redshifts \cite{wol86}.
Since then, much of the modeling has incorporated rotating
disks as a working hypothesis (e.g., \cite{sch90,fal93,kau96}). 
The first attempt to infer
the velocity field of the gas from the data was made by
Briggs {\etal} (1985) who argued the red-blue asymmetry 
detected in the metal lines of a 
$z$ $<$ 1  damped absorber could arise from 
passage of the line of sight through a rotating disk. 
Working with higher resolution data
Lanzetta \& Bowen (1992) presented formulae for
velocity profiles arising from highly inclined
sightlines through a rotating 
spheroid, that mimic sightlines through a thin rotating disk,
in order to model the asymmetries they found in
three damped systems with $z$ $<$ 1. 
Using even more accurate HIRES spectra 
Wolfe (1995) noticed similar
asymmetries in the metal-line profiles of five  damped systems with
$z$ $>$ 1.7. Wolfe (1995) assumed the velocity profiles
arose in rotating disks of gas in which the gas distribution
decreases exponentially both with radius and vertical distance
from the plane.
A Monte Carlo approach to this problem  was outlined,
and a preliminary description of the present
work was given in Wolfe (1996).

In $\S6$ the techniques described in this section will be
applied to: (1) rapidly rotating ``cold'' disks of massive
galaxies; (2) slowly rotating ``hot'' disks of dwarf galaxies;
and (3) rotating disk progenitors of ordinary spirals
predicted by the CDM cosmology.
 
\subsection{Kinematic Model}

Assume the velocity profiles in damped 
{\lya} systems are
explained by passage of the line
of sight through a centrifugally supported disk
in which the volume density of neutral gas
is given by

\begin{equation}
n(R,Z) = n_0 \exp \ltk - {R \over R_d} - {|Z| \over h}  \rtk \;\;\; ,
\label{TRDvol}
\end{equation}

\noindent where $R$ and $Z$ are
cylindrical radius and vertical displacement from midplane, 
$n_0$ is the central gas density,
$R_{d}$ is the radial scale length, and $h$ is the vertical
scale height.  The column density of gas perpendicular to the disk
is thus

\begin{equation}
N_\perp (R) = \Nperp \exp \ltp -R / R_d \rtp \cmma
\label{Ncolm}
\end{equation}

\noindent where the central perpendicular column density,

\begin{equation}
\Nperp = 2 n_0 h.
\label{Nperpdef}
\end{equation}

\noindent Because the velocity profiles in Figure~\ref{sptra}
indicate the density field comprises a discrete rather than
continuous distribution of gas in velocity space, 
$n(R,Z)$ is proportional to the number of discrete clouds (i.e.,
velocity components) per 
unit volume, if, as we assume, the column density per cloud remains
constant.  Further suppose the gas rotates with a flat rotation curve
given by

\begin{equation}
v_{\phi}(R,Z) = v_{rot},
\label{vroteq}
\end{equation}

\noindent where $v_{\phi}$ is the azimuthal component of the
velocity.
Although equation~\ref{vroteq} breaks down at values of $Z$ 
where
the $R^{th}$ component of the gravitational force 
differs significantly from its value at
$Z$ = 0, we shall assume this equality is
valid at all $Z$, and re-examine this assumption in $\S$ 9.  
In addition, let the clouds exhibit isotropic random motions
characterized by a one-dimensional velocity dispersion,
$\sigma_{cc}$.

\begin{figure}
\centering
\includegraphics[height=4.0in, width=3.0in]{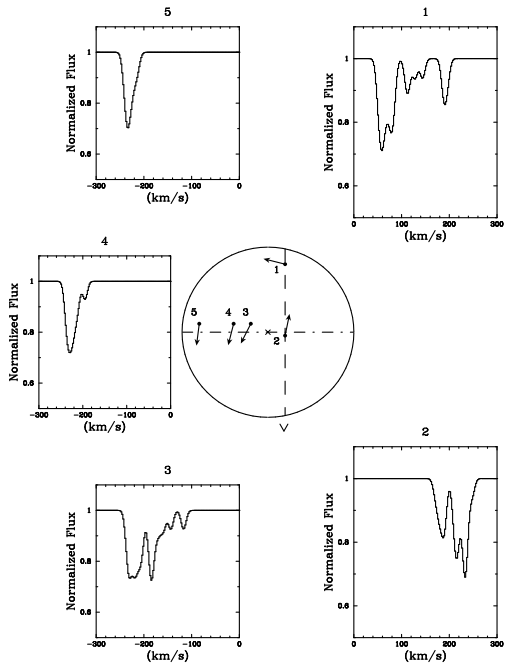}
\caption{The center circle represents an inclined disk (top view)
rotating counter clockwise with rotation speed $v_{rot} = 250$ \kms.
The solid dots represent the intersection points for 5 different
sightlines with the midplane of the disk.  
The solid arrows indicate the direction of the rotation vector.  
The dashed vertical line represents the 
sightline for intersections 1 and 2.  
Finally, the dash-dot line
is the kinematic major axis, i.e. the intersection of the plane
of the disk and the plane of the sky.
The sightlines are inclined by
70 $\deg$ with respect to the normal of the disk and yield the profiles
labeled 1$-$5. See the text for a discussion of the profiles. }
\label{mlos}
\end{figure}

The profile asymmetry is the product of two effects. The ``radial effect''
stems from the $R$ dependence of $n(R,Z)$ {\em along the line of sight}.
This produces a cloud distribution  symmetric about the
peak located where $R$ equals the impact parameter, $b$, i.e., where the
sightline crosses the major axis of the inclined disk (see Figure~\ref{mlos}
where the major axis is denoted by the horizontal 
dash-dot line and the line of sight
by the vertical dashed line).  In this case
the strongest absorption occurs
at the smallest value of $R$ where the magnitude of the rotational velocity
projected along the line of sight is largest; i.e., the ``tangent
point'' used in 21 cm investigations of the Galactic rotation
curve \cite{mihalas81}.
As $R$ increases, the absorption
gets weaker and the magnitude of the projected rotational velocity decreases,
thereby producing a velocity profile with strong absorption
at one side of the profile and progressively weaker absorption
toward the other.
The ``perpendicular effect'' stems from the $Z$ dependence
of $n(R,Z)$ along the line of sight. If the line of sight
intersects midplane, $Z$ = 0, more than a few scale heights
from the major axis, the cloud distribution will peak at the
intersection point rather than the major axis, 
because $h$ $<$ $R_{d}$. The resultant cloud
distribution along the line of sight is lopsided with
more clouds expected toward than away from the major axis,
owing to the ``radial effect''.  The asymmetry of this profile is
the reverse of the previous profile, 
because the absorption is strongest where the magnitude of
$v_{rot}$ projected along the line of sight is smallest and weakens as
the magnitude of this velocity component increases toward the major axis.

Figure~\ref{mlos} shows examples of velocity profiles arising
from these effects. The profiles designated 1 and 2 are
caused by sightlines intersecting midplane at locations
1 and 2. The profiles
are produced by absorption in discrete clouds and are
computed according to the prescription given in the next subsection.
The asymmetry of profile 2 is dominated by the radial effect
because the midplane intersection is so close to the major
axis.  Profile 1
exhibits the  reverse asymmetry because the midplane intersection
is so far from the major axis.
Figure~\ref{mlos} also shows how the velocity profiles are
affected by changes in $b$. 
Profiles 3$-$5 arise at midplane intersections with increasing $b$
which are displaced from the major axis by the same distance along
the line of sight.
The absorption intervals, $\Delta v$, decrease with
increasing $b$, because the gradient of $v_{rot}$ projected along the
line of sight decreases with increasing $R$. 

\subsection{Monte Carlo Methods}
\label{MCmthds}

To generate a statistical ensemble of velocity profiles we adopt a Monte Carlo
technique in which sightlines penetrate the midplanes of randomly
oriented disks at random locations. The clouds responsible for 
absorption are randomly selected from the $n(R,Z)$ distribution
constrained at fixed impact parameter, $b$ = $R$cos$\phi$,
where $\phi$ is the azimuthal angle in the plane of the disk.
More specifically, we use the following prescription:

(1) Pick input parameters $N_{\perp}(0)$, $N_c$, $\sigma_{int}$,
$h/R_{d}$, $\sigma_{cc}$, and $v_{rot}$,
where $N_{c}$ is the H I column density
per cloud, and $\sigma_{int}$ is the
velocity dispersion internal to a cloud. For all our disk models we let
$\sigma_{int}$ = 4.3 {\kms}. 
The value of $\sigma_{int}$  approximates the internal
velocity dispersions of Gaussian velocity components used to model
velocity profiles in damped systems \cite{wol94,pro96,pro97}. 
Because the velocity profiles do not depend
on an absolute length, we express all length scales 
in units of $R_{d}$. 
The remaining parameters either remain fixed during the simulation of
a given ensemble of velocity profiles, i.e, during one run, 
or vary from one profile to the next.

(2) Randomly select the angle of inclination $i$ of the disk, where $i$ is
the angle between the normal to the disk and the line of sight.

(3) Randomly draw coordinates of the midplane intersection, 
($X,Y$), from the uniform projection of
the inclined disk onto the plane of the sky, i.e., onto the ellipse

\begin{equation}
{X^2 \over R_f^2} + {Y^2 \over R_f^2 \cos^2 i} \leq 1 \;\;\; ,
\label{felps}
\end{equation}

\noindent where the $X$ axis coincides with the major axis of the inclined
disk.  We set $R_f$ to be
sufficiently large that sightlines outside
the ellipse will {\em not}
encounter $\N{{\rm H I}} \geq 2 \sci{20} \cm2$.  For the values of $\Nperp$
adopted here, we find that $R_f \approx 10 \, R_d$ is
satisfactory.  Given the midplane intersection and inclination angle,
the line of sight is uniquely defined by setting $X$ equal to the
impact parameter,  $b$.

(4) Compute $N(X,Y)$, the H I column density along the 
line of sight, from the integral $\intl_{X=b} n(R,Z)ds$ where $ds$ is the 
differential length element along
the line of sight and $R$ = $R(s)$ and $Z$ = $Z(s)$. 

(5) Choose a minimum value for $n_c$ at the $\N{HI}$ threshold (see below)
and then for columns above threshold let $n_c$ be proportional
to  $N (X,Y)$. 
After $n_{c}$ is selected, pick clouds randomly along the line of sight
according to the distribution $n(R,Z)$.
 
(6) Determine the ionic column density 
and oscillator strength, $f$, that gives rise to the
synthetic profile for the adopted transition with wavelength
$\lambda$. We assume
$\lambda = 1808.0126 \rAA$, the wavelength of the Si II 1808
resonance transition, $10^{12.5} \cm2$ for the ionic
column density of each cloud, and normalize $f$ such that the product of
$N_c f \lambda$ produces a line profile that satisfies
the line selection criteria, i.e. having the observed 
$\bar\tau(v_{pk}) / \sigma (\bar\tau)$.
Therefore, the choice of $N_c$ and $\lambda$ are arbitrary.

(7) Compute the simulated profile by (a) solving the transfer
equation for the cloud ensemble, (b) smoothing the spectrum to the
instrumental resolution of HIRES, and (c) adding Gaussian noise
such that the SNR of the synthetic profiles corresponds to
each of the {\ndmp} observed profiles in Figure~\ref{sptra}.
Repeat this procedure 500 times for
each observed profile in order to compute 
{\nrun} synthetic velocity profiles for a given run.

Let us discuss the input parameters in more detail.
The profiles are obviously sensitive to $v_{rot}$, $h/R_{d}$,
and $\sigma_{cc}$/$v_{rot}$. 
In the limit $\sigma_{cc} \ll v_{rot}$,  the velocity widths
are mainly affected by $v_{rot}$ and $h/R_{d}$,
because $\Delta v$ $\propto$ $v_{rot}$ and a fixed length
of sightline samples an increasing fraction of the rotation
curve as $h/R_{d}$ increases. The symmetry of the profiles
is affected by $\sigma_{cc}/v_{rot}$ and $h/R_d$, because the systematic
nature of the velocity field is diluted as the contribution
made by random motions increases and because small values of 
$h/R_d$ result in narrow symmetric profiles. The profiles 
are indirectly sensitive to 
$N_{\perp}(0)$.  The portion of the disk
selected for the statistical sample lies within the elliptical
contour defined by $N(X,Y)$ $\ge$ 2$\times$10$^{20}$ cm$^{-2}$, 
hence the quantity $N_{\perp}(0)$ determines the
size of the ``available'' disk. If
$N_{\perp}(0)$ is large, the ``available'' disk extends out to many
radial scalelengths, $R_{d}$. In that case the impact parameter to a typical
sightline will be large compared to $R_{d}$. This produces a
small gradient in 
$v_{rot}$ projected along the line of sight, that results in
profiles with small {$\delv$} (see sightlines 3$-$5 in Figure~\ref{mlos}).
Conversely when $N_{\perp}(0)$ is small, larger $\Delta v$ will
be typical. 
We shall refer to this effect as the ``column-density kinematic effect''.
In order to reproduce the observed distribution
of H I column densities, we selected $N_{\perp}(0)$ 
with values between 10$^{20.8}$ and 10$^{21.6}$ cm$^{-2}$. 

\begin{figure}
\centering
\includegraphics[height=6.5in,width=3.5in,angle=-90]{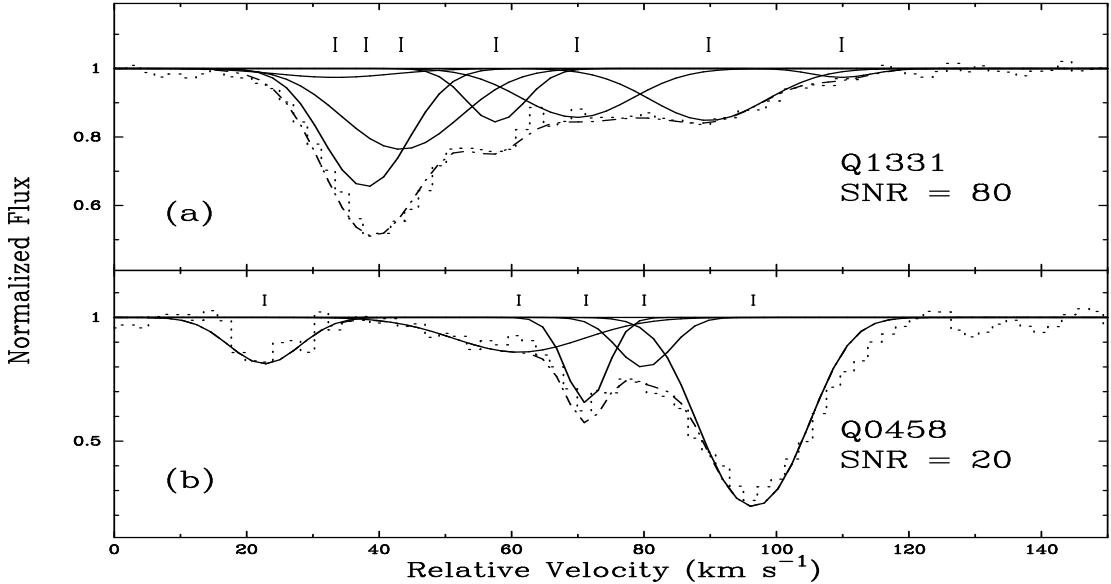}
\caption{Comparison of two line-profile fits to (a) the Si II 1808
transition from the $z=1.776$ damped system toward Q1331$+$170 and
(b) the Cr II 2056 transition from the $z=2.040$ system toward
Q0458$-$020.  In both figures, the dotted histogram is the observed
profile, the solid lines represent the individual components of the 
VPFIT solution, and the dash-dot line is the convolved final solution.
Note that even though profile (a) has a smaller $\delv$ than that of
the Q0458 profile,
the VPFIT solution required 2 more clouds than the lower SNR Q0458 data.}
\label{fig1331}
\end{figure}

On the other hand the velocity profiles are much less sensitive
to $n_{c}$, the number of clouds along the line of sight,
provided $v_{rot}$ $>>$ $\sigma_{cc}$; i.e., when the kinematics
are dominated by systematic rather than random motions.
The decomposition of the
absorbing gas into multiple discrete clouds is based on our experience
with fitting model profiles to the high-resolution HIRES spectra
(see \cite{wol94,pro96,pro97}).
Using chi square fitting techniques we find that $\approx$ five 
Gaussian velocity components 
are the {\em minimum} number necessary for acceptable fits, with
no strict upper limit. That is, 
we find trials with $n_c = 60$
yield test statistics ($\S$~\ref{ststs}) similar
to those trials with $n_c \approx 5$. 
Although the narrowest 4 profiles in the observed
data set can be modeled with only 1 or 2 components, 
these features are probably blends of a greater number ($> 5$) of components
superimposed in velocity space.  
The case for large numbers of components is supported by the
fact that when $\Delta v$ is fixed, $n_{c}$ increases with increasing SNR.
In Figure~\ref{fig1331} we show our 
model fits to (a) the Si II 1808 profile for the $z=1.776$ damped
system toward Q1331$+$170 and (b) the Cr II 2056 profile for
the $z=2.040$ system toward Q0458$-$020.  Although $\delv$ for
the Q1331 profile is below the mean
for the sample, our best solution requires $n_c = 7$.  By contrast
we need only 5 components to fit the wider, but noisier Cr II
profile from the $z=2.04$ damped system toward Q0458$-$020.

Therefore, in each run we adopt $n_{c}$ = 5 for threshold impacts,
$N(X,Y) = 2 \sci{20} \cm2$, and then 
let $n_c$ be proportional to higher values of  $N(X,Y)$ for
impacts within the threshold ellipse. The crucial point
is that the results are insensitive to $n_{c}$ for $n_{c}$ $>$ 5.
The minimum value for $n_c$ is important, because if 
it were significantly smaller, velocity asymmetries could
arise from strong Poissonian fluctuations in the strengths of the
velocity components rather than from gradients in
systematic velocity fields. 
We also find the results discussed below 
to be insensitive to $\sigma_{int}$, provided
it is small compared to $v_{rot}$ or $\sigma_{cc}$.

\section{MODELS WITHOUT ROTATING DISKS}

We now discuss the kinematic predictions of other models suggested to
explain the damped {\lya} systems. In particular we consider 
models in which rotation does not dominate the velocity field. 
The Monte Carlo methods used for these tests are essentially
those presented for the disk model in $\S$~\ref{MCmthds}.

\subsection{Isothermal Halos of Normal Galaxies}

	The low metallicities measured in damped
{\lya} systems, [Zn/H] $\approx$ $-$1.2 \cite{ptt94}, and
the pattern of relative metal abundances in these objects 
\cite{lu95b,lu96b} bring to mind  
Galactic halo stars; i.e., stars that were metal enriched
only by type II supernovae. For these reasons Lu {\etal} (1996b) suggest
the damped systems comprise neutral clouds in galactic halos. The abundance
patterns of damped {\lya} systems argue against metal-enriched disks
of current spiral galaxies, because
metal enrichment by type I as well 
as type II supernovae are required  to explain their abundance patterns. 
We wish to test the halo hypothesis by modeling
the cloud kinematics with an isothermal halo.

\begin{figure}
\centering
\includegraphics[height=4in, width=3in]{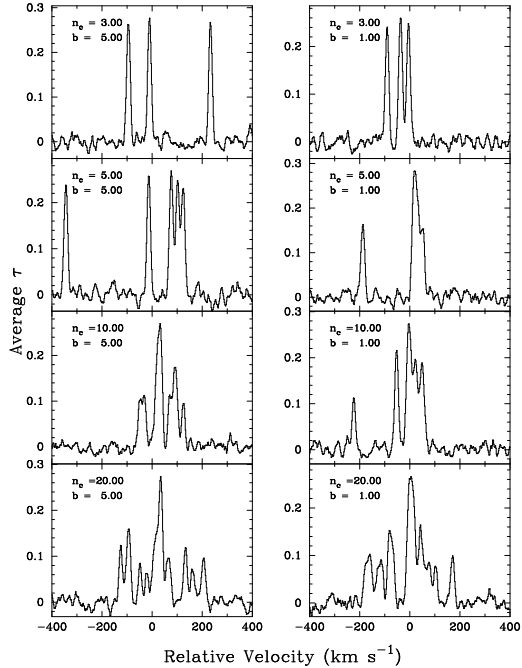}
\caption{Average optical depth profiles derived from the 
isothermal halo model with varying cloud number $n_c$
and impact parameter $b$.  Note the profiles
tend to peak near $v=0$ particularly as $n_c$ increases.  
The profiles also have very large velocity intervals,
exceeding 200 \kms in the majority of cases.}
\label{isofig}
\end{figure}

To determine the velocity profiles we  consider 
an isothermal halo model  in which $n(r)$, the gas density
at 3D radius $r$,
is proportional to the total mass density $\rho(r)$ given
by Table 4-1 in Binney and Tremaine (1987) for self
gravitating isothermal spheres. The proportionality
constant is fixed by setting the universal baryon to dark
matter ratio to 0.1 \cite{nav95}.  As a result, all
the baryonic mass in galaxies is initially placed in the halo.
In this model the mass of  neutral gas 
in the halo cannot vary
significantly between
$z$ $\approx$ 3 and  1.5 in order to explain the relatively constant
comoving density of neutral gas observed in this redshift interval 
\cite{stor97}. Because halos are dominated by random rather
than systematic motions such as rotation, the asymmetries predicted for
the ``cold'' rotating disks should be suppressed in this model,
provided significant velocity gaps between the clouds do not exist.
This is illustrated in Figure~\ref{isofig} 
showing eight $\bar\tau(v)$ profiles generated by a model of this type.

\subsection{Spherical Accretion onto Galaxy Bulges}

The infall of gas onto pre-existing mass concentrations is a plausible
scenario for galaxy formation, one that 
may give rise to the velocity profiles
in damped {\lya} systems. Such mass 
concentrations may have recently been discovered
in the form of compact galaxy bulges with redshifts of $z \approx$ 3.5
\cite{ste96a}.  The regularity of the 
brightness distributions further suggests these objects 
formed at even higher redshifts,
and reinforces the idea that the remaining parts of the galaxy
may have formed by infall of neutral gas onto a pre-existing bulge component
\cite{pbl93}.  Gaseous infall was first suggested by 
Arons (1972) who proposed that QSO absorption lines originated
in gas partaking in the gravitational collapse of low mass
protogalaxies. Because he 
focused on the highly ionized gas   
responsible for the low column-density {\lya} forest systems,
his model is not directly applicable to the 
neutral gas comprising the damped {\lya} systems. 

\begin{figure}
\centering
\includegraphics[height=4in, width=3in]{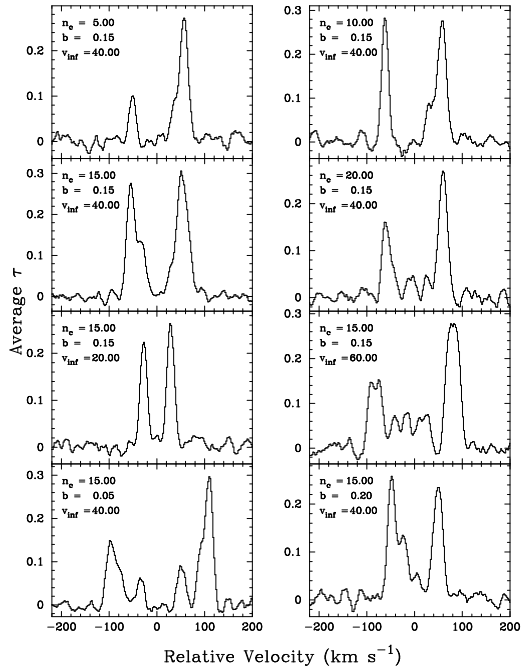}
\caption{Average optical depth profiles derived from the 
spherical accretion model with varying cloud number $n_c$,
impact parameter $b$, and infall velocity $v_{inf}$.  Unlike
all of the other models considered, these profiles exhibit peaks
at both edges of the velocity interval.}
\label{sphfig}
\end{figure}

We model the infall by assuming neutral clouds responsible for 
the damped {\lya} systems are embedded in a hydrodynamical infall
of fluid onto a spherical mass distribution. By assuming 
spherical symmetry we can adopt
the analysis of Bondi (e.g. \cite{shp83}) for 
hydrodynamical spherical accretion. In this case 
the radial velocity
and volume density of the fluid are given by

\begin{equation}
v(r) = -v_0 \ltp {r_s \over r} \rtp^\ohf \;\;\;\; [r \ll r_s]
\;\;\; ,
\end{equation}

\noindent and

\begin{equation}
n(r) = n_0 \ltp {r_s \over r} \rtp^\rhf \;\;\;\; [r \ll r_s]
\;\;\; , 
\end{equation}

\noindent where $v_0$ and $n_0$
are the radial velocity and  density
at $r = r_s$. The quantity $r_s$ is the transonic radius and is related to the 
ambient thermal energy per unit mass (i.e., the sound speed squared), $a_\infty^2$, by

\begin{equation}
r_s \approx {G M \over a_\infty^2} \perd
\end{equation}

\noindent In this model the gas
has zero net angular momentum;  future work will address a scenario
with finite net angular momentum.

Figure~\ref{sphfig} shows
examples of velocity profiles produced by a typical accretion model. 
Notice
the double absorption peaks 
located at both edges of the profile. These are caused by
the maxima in the line-of-sight component of $v(r)$ at
displacements $\pm$ $b$ from the
point of closest approach. 
By contrast the profiles 
produced by ``cold'' rotating
disks tend to produce single peaks at the profile edges
(see Figure~\ref{mlos}). 

\subsection{Random Motion Models}
\label{ranm}

In order to generate a control sample, that is, a sample of
velocity profiles created by velocity fields without
inherent systematics, we investigate a model in which the
velocities are randomly drawn from a uniform velocity distribution 
covering the velocity interval $v$ = [0,$v_{max}$].
The model has a generic problem matching the observations
in that it yields nearly the same $\Delta v$ for
each simulated line profile.
This is in obvious conflict with the observed profiles 
exhibiting a wide range of $\Delta v$.
Because the  neutral volume density, $n$, is not described by
this model, it does not yield values of $N$(H I).  
Therefore, for each run we arbitrarily select the number of individual
clouds from the interval $n_{c}$ = [5,25],
resulting in variations of $N$(H I).

\section{STATISTICAL TESTS}
\label{ststs}

\begin{figure}
\centering
\includegraphics[height=4in, width=3in]{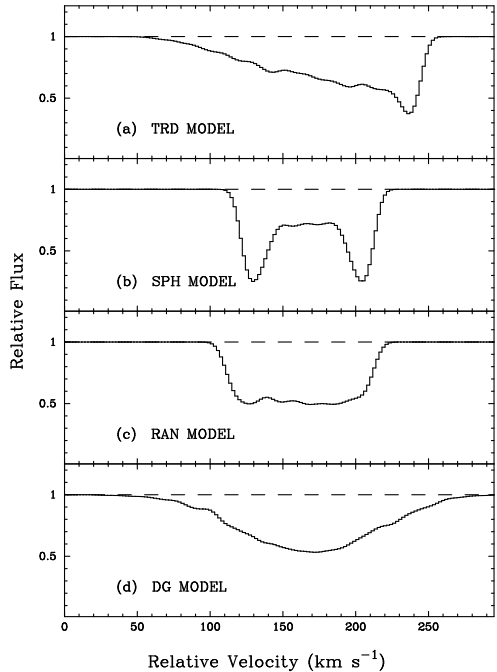}
\caption{Smooth absorption profiles for the (a) Thick Rotating Disk (TRD) 
(b) Spherical Accretion (SPH) (c) Random Motion (RAN) and (d) Dwarf Galaxy
(DG). Profiles derived by setting $n_c = 1000$.}
\label{smooth}
\end{figure}

Figure~\ref{smooth} is a plot of characteristic line profiles for
each of the models for a smooth distribution of gas ($n_{c} \gg 1$).
Rapidly rotating disks are characterized by 
asymmetric distributions in optical depth
that tend to peak at one edge of the velocity interval.
The spherical accretion model produces peaks
at both edges, the random motions model produces a flat profile without
peaks, while central peaks or null peaks
characterize ``hot'' models where $\sigma_{cc}$ $\gtrsim$ $v_{rot}$. 
In this section we design test statistics to discriminate
among these models.  In particular,
the tests focus on the width of the features in velocity space
(the velocity interval),
the location of the peak component within the velocity interval,
and the degree of asymmetry evident in the profiles.

\subsection{Velocity Interval Test}
\label{velts}

The velocity interval, $\delv$, of an absorption profile is defined
as the width of the profile in velocity space. 
A simple method for determining $\delv$
is to identify the $3 \sigma$ components
at the lowest and highest velocities 
of the ${\bar \tau(v)}$ array 
and define  $\Delta v$ as the difference
in velocity between these components.  This approach, however,
gives weak outlying components the same weight as strong
components in the main absorption complex. As a result
a single, weak component displaced 200 {\kms} from the centroid
of a strong absorption complex 
spanning a velocity interval of 40 {\kms} 
would yield $\Delta v \approx$ 220 {\kms} even though most of the
absorption spans a much smaller $\Delta v$.

For this reason we define
$\Delta v$ as the velocity interval
weighted by optical depth. Specifically, 
define $\Delta v$ by first computing the integrated
optical depth

\begin{equation}
 \tau_{tot} = \int {\tau(v)}dv \perd
\end{equation}

\begin{figure}
\centering
\includegraphics[height=4in, width=3in]{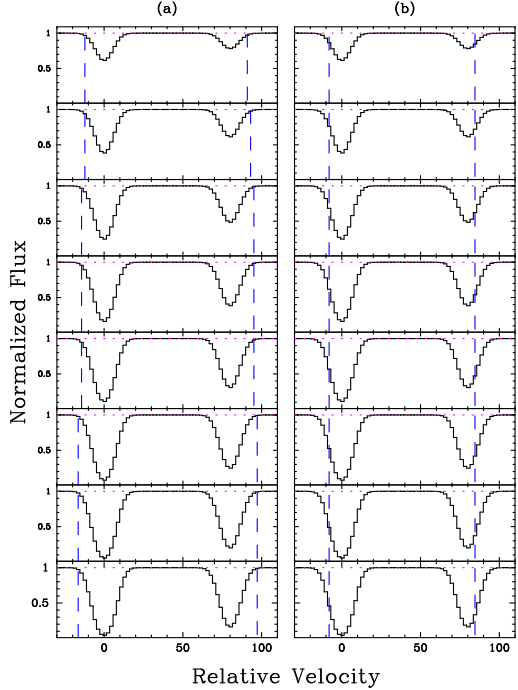}
\caption{Comparison of the effects of saturation on the definition
of the velocity interval by (a) a 3$\sigma$ limit in the $\bar\tau(v)$
array and 
(b) designating 90$\%$ of the total optical depth.  
Note that the velocity interval for the
optical depth method remains constant with increasing line strength, while
the velocity interval derived with the 3$\sigma$ limit clearly
increases ($\approx 20\%$ from the first to last frames).}
\label{fvint}
\end{figure}

\noindent We then step
inward from each edge of the profile until reaching the pixel
where 5$\%$ of
$\tau_{tot}$ is removed, and  set $\Delta v$ equal to the
velocity difference between the pixels.
This definition removes outlying weak components
that may be statistical fluctuations that are
not good tracers of the underlying
velocity field. While the routine may underestimate the true
velocity interval,
this is preferable to including misleading statistical artifacts.
In addition, numerical tests (see Figure~\ref{fvint}) reveal 
the removal algorithm is not sensitive to the optical depth
of the absorption feature.  Although the choice of removing
$10\%$ of $\tau_{tot}$  is somewhat arbitrary, 
the simulations indicate the other statistical tests are largely
independent of the removed fraction of $\tau_{tot}$.  

\subsection{Mean$-$Median Test}

One of the principal characteristics of the velocity profiles
is their asymmetry, and so we devise a number of test statistics
to discriminate asymmetric profiles from symmetric ones.
The Mean$-$Median Test gauges 
asymmetry by measuring the relative difference between
the mean and median velocities of the
optical depth profile.  The test first establishes
the velocity interval as discussed above.  The mean velocity is
defined as the velocity of the midpoint of the interval where the 
edge corresponding to the smallest velocity
is normalized to $v=0$ (i.e. $v_{mean} = \ohf \delv$).  
The routine
calculates $\tau_{tot}$ within the velocity interval, and then
determines the median velocity, $v_{median}$, 
as the velocity  dividing $\tau_{tot}$
in half.  The formal statistic is expressed as:

\begin{equation}
f_{mm} = {| v_{mean} - v_{median} | \over (\delv / 2)} \;\;\; .
\end{equation}

\noindent  For a perfectly symmetric distribution of gas, $f_{mm} = 0$,
while more asymmetric profiles will tend to $f_{mm} = 1$.  The 
test is both physically motivated and relatively simple.  
Note that because the Velocity Interval routine 
removes $10\%$ of $\tau_{tot}$ symmetrically, it does not affect
the Mean$-$Median Test.

\subsection{Edge-Leading Test}
\label{edgtst}

The test statistic, $f_{edg}$,  measures the relative velocity difference 
between $v_{pk}$, the velocity of the feature with peak ${\bar \tau}(v)$,
and $v_{mean}$ such that 

\begin{equation}
f_{edg} = {| v_{pk} - v_{mean} | \over ( \Delta v / 2)} \;\;\; .
\end{equation}

\noindent Thus, $f_{edg}$ = 0 for peaks at the center of the profile, 
while very ``edge-leading'' profiles yield $f_{edg} \approx 1$.

\subsection{Two Peak Test}

 A third test statistic, 
designed to accentuate the location of the second strongest
peak in the velocity profiles, is given by

\begin{equation}
f_{2pk} = {\pm}\frac{\vert v_{2pk}- v_{mean} \vert}{(\Delta v/2)},
\end{equation}

\noindent where $v_{2pk}$ is the velocity of the second strongest peak.
The plus sign holds if the second peak is between the velocity
of the first peak, $v_{pk}$,  and $v_{mean}$, otherwise the 
negative sign holds. Consequently,
profiles with peaks at {\em both} edges, as would arise
in the case of spherical accretion, or
profiles in which  $v_{2pk}$ $<$ $v_{pk}$,
that might be drawn from Gaussian velocity distributions, can be distinguished from 
the peaks arising from rotating exponential disks.
The disk model tends to produce peaks
with $v_{pk}$ $<$ $v_{2pk}$ $<$ $v_{mean}$
because $\tau(v)$ systematically
decreases with increasing $v-v_{pk}$.
In fact, the Two$-$Peak Test was introduced to differentiate
between the rotating disk and spherical infall models.
In the random models the two strongest peaks
should be uncorrelated and will occur at velocities uniformly
distributed within $\Delta v$. 
While uncorrelated, the two strongest
peaks will be more centrally located
in models for which $\sigma_{cc}$ $>$ $v_{rot}$, since
the random part of the velocities are drawn from a Gaussian
rather than a uniform velocity distribution.

\begin{figure}
\centering
\includegraphics[height=4in, width=3in]{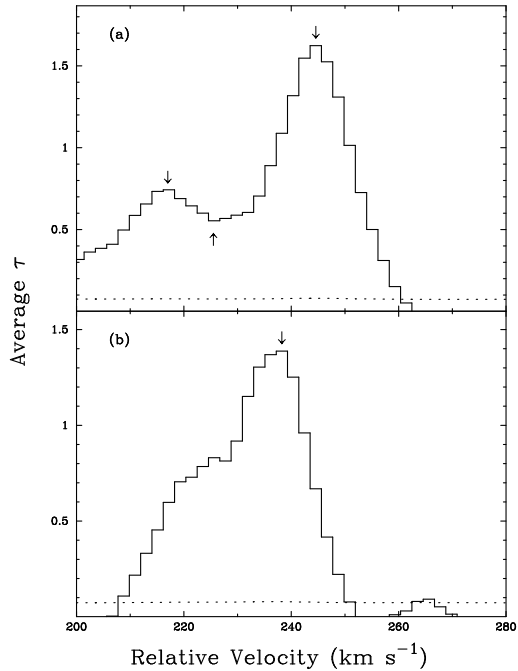}
\caption{Average optical-depth profiles for two systems, 
(a) one exhibiting two peaks and the other (b) only a single peak.  The
down arrows indicate peaks while the up arrow indicates the boundary
between the two peaks.  This boundary also acts as the artificial
continuum level from which the second peak is considered a 3$\sigma$ feature.
The dotted line is the $1 \sigma$ error array for $\bar\tau$.}
\label{sndpk}
\end{figure}

     As in the Edge-Leading Test, this test first 
finds the strongest feature in the velocity
interval.
It then searches across $\Delta v$ for a second
significant feature satisfying two criteria: 
(1) $\bar \tau (v_{2pk}) > {1 \over 3} \bar \tau(v_{pk})$;
and (2) $\bar \tau(v_{2pk})$ is a $3 \sigma$
feature above an artificial continuum level set by the lowest
$\bar\tau(v_i)$ value reached between $v_{pk}$ and $v_{2pk}$.
The first criterion prevents small, physically irrelevant,
features from biasing the test.  The second criterion
is necessary to distinguish a physically distinct second peak from
a small rise on a single feature (see Figure~\ref{sndpk}).  
After the two peaks are determined, the test measures the relative velocity
of the second strongest peak from the center of the velocity interval.
Note that if there is no significant second peak, the value for
$f_{edg}$ is adopted instead.  Therefore, models 
preferentially yielding profiles with single peaks will 
have an $f_{2pk}$ distribution biased to positive values.

\section{RESULTS}
\label{rstssec}

Having generated ensembles of velocity profiles for each of the models
in $\S$ 3 and $\S$ 4  we will use them to construct
distributions of the test statistics
discussed in $\S$ 5. We 
test the validity of the null hypothesis, i.e., that
a given model is compatible with the data, by
comparing model and empirical distributions with a
two-sided  Kolmogorov-Smirnov (KS) test. 
The KS test gives the probability $P_{KS}$
that the two distributions
could have been drawn from the same parent population. Because it is
the most conservative test of its kind, a low value for $P_{KS}$ is
strong evidence that a given model is unlikely to represent the data.
 
\begin{figure}
\centering
\includegraphics[height=8in, width=6.5in]{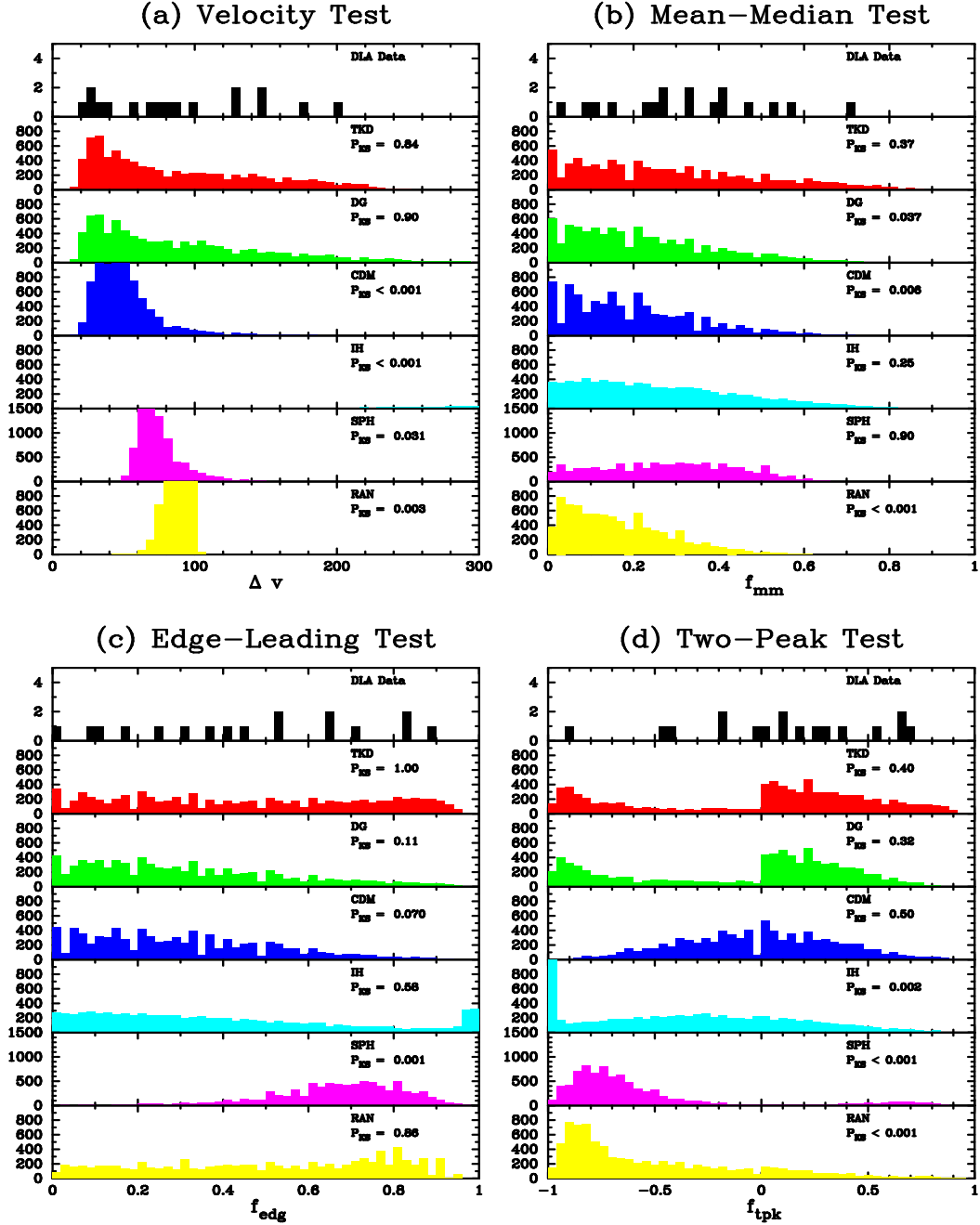}
\caption{Plot of the $\delv$, $\f{edg}$, $\f{mm}$, and $\f{2pk}$
distributions for the data and the six test models.
The $P_{KS}$ values represent the probability the test distribution
and the data could have been drawn from the same parent population.
Note that every model except the TRD Model is inconsistent with
the data for one or more statistical tests.}
\label{fail}
\end{figure}

Figures~\ref{fail}a$-$d show comparison plots 
of the 4 test statistics for each of the 6 kinematic models.
After an exhaustive search of the parameter space $\und{permitted}$ 
by each model, we chose those parameters yielding the closest
agreement to the data.  Table~\ref{param} gives the parameters  
used in the simulations, while the $P_{KS}$ values are listed in 
Table~\ref{ksfail}.  The Figure demonstrates that apart from 
``cold'' disks with high rotation speeds,
the remaining models are ruled out at high
levels of confidence by at least 1 statistical test. 
Therefore, of all the {\em tested} models,
only ``cold'' disks with rapid rotation are
consistent with the damped {\lya} kinematics. 
In the following subsections, we will briefly
discuss why the other models fail specific tests
and why improved results are unlikely.

\begin{table*}
\caption{MODEL PARAMETERS}
\begin{center}
\begin{tabular}{lccccc}
\hline
\hline
MODEL & $\Nperp$
& $v_{rot}$ or $v_0$
& $\sigma_{cc}$
& $n_c$ & $h$ \\
& ($\cm2$) & ({\kms}) & ({\kms})
& $(2 \sci{20})$ & $(r_0)$\\
\hline
TRD$^{a}$ & 21.2 & 250   & 10  & 5  & 0.4 \cr
DG$^{b}$  & 21.2 & 50    & 5-55& 5  & 0.3 \cr
CDM$^{c}$ & 21.2 & 30-200& 10  & 5  & 0.3 \cr
IH$^{d}$  & --   & 40    & 190 & 10 & -- \cr
SPH$^{e}$ & 19.4 & 20    & 10  & 5  & -- \cr
RAN$^{f}$ & --   & 100   & --  & 15 & -- \cr
\end{tabular}
\label{param}

$^{a}$ Cold, Rapidly Rotating Disk Model

$^{b}$ Dwarf Galaxy Model

$^{c}$ Cold Dark Matter Model

$^{d}$ Isothermal Halo Model

$^{e}$ Spherical Accretion Model

$^{f}$ Random Motions Model

\end{center}
\end{table*}

\begin{table*}
\caption{$P_{KS}$ VALUES} \label{ksfail}
\begin{center}
\begin{tabular}{lcccc}
\hline
\hline
MODEL
& VEL
& MM
& EDG
& TRD \cr
\hline
TRD & 0.84            & 0.37           & 0.99   & 0.40           \cr
DG  & 0.90            & 0.034          & 0.11   & 0.31           \cr
CDM & $2.0 \sci{-4}$  & 0.0062         & 0.070  & 0.49           \cr
IH  & $4.5 \sci{-16}$ & 0.25           & 0.58   & 0.0016         \cr
SPH & 0.030           & 0.90           & 0.0015 & $1.0\sci{-8}$  \cr
RAN & 0.0034          & 4.5 $\sci{-4}$ & 0.86   & $4.0 \sci{-6}$ \cr
\end{tabular}
\end{center}

\end{table*}

\subsection{Models with Rotating Disks}

Rotating disks are essential elements in 
models with ``cold'' disks of giant spirals,
``hot'' disks of dwarf galaxies, and
the progenitors of spiral galaxies
in CDM cosmologies.  The models considered in this
subsection have the properties discussed in $\S$ 3, and differ
only in the choice of
the parameters $v_{rot}$, $\sigma_{cc}$, $h/R_d$, $n_c$
and $\Nperp$.

\subsubsection{``Cold'' Disks of Giant Spirals}
\label{raptkd}

In this model, velocity profiles are generated by line-of-sight
impacts through a set of intrinsically {\em identical} exponential disks
that are ``cold'' and rapidly rotating.
Therefore, all the input parameters are the same for each of the
{\nrun} impacts comprising a given run. 
In order for the H I disks to be ``cold'' 
we let $\sigma_{cc}$ = 10 {\kms} which
equals the velocity dispersion measured across the H I disks of
spiral galaxies \cite{van84}. 
Note the results are insensitive to the value of $\sigma_{cc}$ 
provided 
$\sigma_{cc} < 25$ \kms; i.e.,  the disks remain ``cold''.
For larger values of $\sigma_{cc}$, the model fails the
Two-Peak Test and also
cannot reproduce the low $\delv$ tail of the Velocity Interval
Test distribution.  For each run we set $v_{rot}$ = 250 {\kms} and  
$h/R_{d}$ = 0.4. This rotation speed is a representative
value for current giant spirals \cite{rub85}.
The scale-height is chosen to reproduce the observed
distribution of $\Delta v$.  

As this model assumes
identical properties for all intercepting galaxies, 
it is somewhat naive.  A more realistic model would include
a distribution of rotation speeds, thicknesses and 
central column densities.
However, it is valuable for testing the null hypothesis that
giant spirals explain the kinematic state of the damped
{\lya} systems.
The figures show that the null hypothesis cannot be ruled
out for this model, because it produces acceptable values of
$P_{KS}$ for each of the 4 statistical tests.

\subsubsection{``Hot'' Disks of Dwarf Galaxies}

In this model we assume the velocity profiles are generated
by the disks of dwarf galaxies in which $\sigma_{cc}$ is
assumed to vary from object to object. 
York {\etal} (1986) first suggested dwarf galaxies
were responsible for QSO absorption systems exhibiting metal
lines. More recently, Pettini {\etal}  (1994) and 
Lu {\etal} (1996b) advocated dwarfs as the 
site of damped {\lya} absorption. These authors were struck
by the similarity between the low metallicities and element abundance
patterns in dwarfs
and damped {\lya} systems.

The velocity profiles are generated by the same algorithms
used to model the ``cold'' disks discussed in the previous
subsection. That is, except for $\sigma_{cc}$,
all the input parameters are fixed for
the profiles computed in each of {\nrun} galaxy impacts.
However, for these disks we 
assume $v_{rot}$ = 50 {\kms} which is a good representation
for the rotation speeds of dwarf irregulars \cite{fre93}. 
We draw $\sigma_{cc}$ from a distribution spanning velocities
between 5 and 55 {\kms} in such a way that the resultant
$\Delta v$ matches the data. The range in $\sigma_{cc}$ simulates the
episodic turbulence originating from the high
frequency of supernova remnants that might occur at large redshifts.
The idea is to let random motions in gas accelerated by
supernova activity produce the observed $\Delta v$.
This was suggested
earlier as the explanation behind the large $\Delta v$
detected in metal line absorbers as well as damped {\lya}
systems \cite{yor86,ptt94}.  Of course, if we were to require
$\sigma_{cc} \ll 55$ \kms always (as observed in local gas-rich
dwarfs), the model fails {\it a priori} as it cannot reproduce
the velocity widths, $\delv > 100$ \kms.
Finally, to further enhance $\Delta v$ we set $h/R_{d}$ = 0.3.

Because the kinematics of the ``hot'' dwarf galaxy
model are influenced by 
velocities randomly drawn from a Gaussian distribution,  
the simulated profiles tend to peak symmetrically about
$v = 0$ \kms.  As a result the model profiles yield discrepant 
$\f{mm}$ statistics; $P_{KS} (\f{mm}) = 0.037$.  
Furthermore, the model is only marginally consistent with the
Edge-Leading Test as $P_{KS}(f_{edg})$ equals 0.11.  
Note that the model survives the Two-Peak Test, because
a large fraction of the profiles are produced with a single
peak, which as noted above, biases the distribution to positive
values in the $f_{2pk}$ distribution.

We wish to point out that these results are significantly dependent
on both the assumed cloud number $n_c$ at threshold $\N{HI}$
(taken to equal 5)
and $\sigma_{cc}$.  Specifically, increasing $n_c$ while holding 
$\sigma_{cc}$ constant will increase the range of $\delv$ values
due to the greater sampling of the random Gaussian distribution.
Similarly, increasing $\sigma_{cc}$ while holding $n_c$ constant
has the same effect.
At the same time, increasing $n_c$ reduces the level of asymmetry
as the random motions begin to dominate the rotation kinematics.
We find that the threshold $n_c = 5$ 
gives the best agreement to all four statistical tests, and
that models with threshold
$n_c > 10$ are ruled out at higher confidence levels,
because both $P_{KS} (\f{mm})$
and $P_{KS} (\f{edg})$ would be less than 0.01.

\subsubsection{Disk Progenitors in CDM Cosmology}

Kauffmann (1996) recently combined Press-Schecter theory
with the Cold Dark Matter cosmology (CDM) to compute the hierarchical
merging of protogalactic disks embedded in dark-matter halos.
Semi-analytic techniques
were used to calculate the formation of 

\noindent H I disks as gas cools and
accretes in the merging halos. With the damped {\lya}
systems in mind, Kauffmann (1996) computed $P(v_{rot})$, the interception 
probability presented by the cross-sectional area of the disks,
as a function of $v_{rot}$  for
various redshifts. The radial distribution of H I 
column densities, $N_{\perp}(R)$ was also computed for different
values of $v_{rot}$.

We calculated velocity profiles for this model as follows.
For each of the {\nrun} impacts comprising a given run
we drew $v_{rot}$ randomly from 
the $P(v_{rot})$ corresponding to the redshift $z$ = 2.5,
in good agreement with the average redshift, $z$ = 2.48, 
of the profiles in our statistical sample. We then fitted exponential
profiles to the plots of $N_{\perp}(R)$ by assuming 
$\Nperp = 10^{21.2} \cm2$.  Kauffmann (1996)
implicitly assumed the disks to be thin, $h$ $<<$ $R_{d}$,
and ``cold'', $\sigma_{cc}$ $<<$ $v_{rot}$. Because
$v_{rot}$ is  less than 100 {\kms} in 70$\%$ of the
model disks, the predicted distribution of $\Delta v$
will not match the observed distribution which extends out
to 200 {\kms}. Therefore, we assumed 
$h$ = 0.3$R_{d}$ to maximize $\Delta v$ without
significantly changing the disk-like character of the model. Otherwise,
we followed the same procedures used to compute velocity
profiles for the other disk models.

The results shown in Figure~\ref{fail} indicate this model is
highly unlikely to represent the data. Owing to the predominance
of slowly rotating disks, 
the Velocity-Interval and Mean-Median Tests result in
$P_{KS}({\Delta v})$ $<$ 0.001 and 
$P_{KS}(f_{mm})$ = 0.002 which are unacceptable. 
Had we increased $\Nperp$ to $10^{22.2} \cm2$, which 
is closer to the value Kauffmann derived, the $\delv$
distribution would peak at even lower velocities, owing to 
the  ``column-density kinematic'' effect, and could be ruled out
at even higher confidence levels.  
{\em Consequently we conclude this CDM
model is ruled out at high levels of confidence.}  Even if we
were to ignore the two highest $\delv$ values, this CDM model
is still ruled out at $> 99\%$ c.l.
The implications of this important
result for theories of galaxy formation are discussed in $\S$ 9.

\subsection{Models Without Rotating Disks}

We now turn to models without rotating disks. 

\subsubsection{Isothermal Halos}

In order to mimic the kinematics of the
Galaxy halo, we let the halo gas rotate with
$v_{rot}(r) = 50$ {\kms}, and assume $\sigma_{cc} = 190$ {\kms}
(see Majewski 1993). As in $\S$~\ref{MCmthds}
we set $\sigma_{int}$ = 4.3 {\kms}. 
For the isothermal sphere we choose the central mass density, $\rho_0$,
in such a way that the mass $M$($r$=20 kpc)
= 2$\times$10$^{11}$$M_{\odot}$, which is  
the mass inferred from a constant {\em disk}
rotation speed $v_{rot}$ = 220 {\kms} out to
$\approx 20 \; \rm{kpc}$.
The length scale of the density distribution, i.e., 
the King radius, is then determined by the relation, $r_{0}$ =
$\sqrt{9{\sigma_{cc}^{2}}/(4{\pi}G{\rho_{0}})}$ and equals 18 
kpc.  However, because the kinematics are dominated by
the random motions of the clouds, the results are essentially independent
of $\rho_0, r_0$, and the assumed density profile.

To generate simulated spectra we follow the 
prescription of $\S$~\ref{MCmthds},
except that the $(X,Y)$ coordinates are drawn from a projected circle
rather than a projected ellipse. The circle has a radius
$r$ = $R_{f}$, where $R_{f}$ has the same 
meaning as in $\S$~\ref{MCmthds}. We
also assume the angular momentum of the gas defines a preferred plane whose
rotation axis is a random variable. 

Figure~\ref{fail}a shows that the value of $\sigma_{cc}$ is sufficiently 
large to throw most of the distribution of $\delv$ off scale, peaking
at $\delv \approx 600$ \kms. As a result the model fails the 
$\Delta v$ test at a very high level of confidence. 
Furthermore, $P_{KS} (\f{2pk}) < 0.002$ owing to the large fraction
of $\f{2pk} < 0$ resulting from the random locations of the
absorption peaks.  We tried to decrease
$\delv$ by reducing $n_{c}$ from 10 to 5, but this increased the
fraction of profiles with $\f{2pk} < 0$, and consequently produced an even lower
$P_{KS} (\f{2pk})$ value.  Because $\sigma_{cc}$
$>>$ $v_{rot}$, this model, like the dwarf galaxy model, generates
profiles exhibiting a high degree of symmetry in the limit $n_c \gg 1$.
However, in this simulation $\sigma_{cc}$ is so large that for
$n_c = 10$, it is rare for two clouds to overlap in velocity space.
Therefore, the strongest absorption peak occurs randomly due to
the noise in the spectrum.  Because this leads to an $\f{edg}$ distribution
similar to that of the Random Motions Model, 
$P_{KS} (\f{edg}) = 0.58$.  By the same reasoning, the asymmetry
is accentuated such that $P_{KS} (\f{mm}) = 0.25$. However,  
we find that for $n_c > 15$, all Isothermal Halo Models are ruled
out by the Mean-Median and Edge-Leading Tests.

\subsubsection{Spherical Accretion}

We modeled the accretion flow with  a central bulge
having  mass $M \approx 10^{10} M_{\odot}$ and gas with
sound speed  $v_0 \approx a_\infty$ = 20 {\kms} 
implying $r_s \approx$ 110 kpc. 
The sound speed is physically reasonable and is chosen to
roughly reproduce the
observed $\Delta v$ distribution.   
However, it corresponds to thermal temperatures
high enough that the gas will be collisionally ionized.
Thus the gas creating damped {\lya} absorption must be
discrete neutral clouds embedded in a smooth, ionized
medium.

By contrast with the other models 
in this section, the kinematics of the spherical infall model
are dominated by systematic motions, i.e. radial infall,
rather than random motions.  
However, the profiles are still symmetric with
the peaks tending to lie at both edges of the velocity interval.
As a result, the Two-Peak Test distribution is
dominated by negative $f_{2pk}$ for all values of $n_{c}$, $v_{0}$
and $n_0$, in contradiction to the empirical $f_{2pk}$ distribution
which is why $P_{KS} (\f{2pk}) < 0.001$.
Furthermore, while the model predicts a peak
in the $f_{edg}$ distribution near $f_{edg}$ $\approx$ 0.75,
the empirical distribution is more uniform between 0 and 1.0.
Thus the profiles predicted by the spherical infall model
are actually too ``edge-leading'', and as a result $P_{KS} (\f{edg}) = 0.001$.  
Note, in order
to reproduce the observed $\Delta v$ distribution
we developed {\em ad hoc} distributions of $v_{0}$.
The combined weight of the tests rule
out this model at very high confidence levels.

\subsubsection{Random Motion Models}

The Random Motion Model assumes a uniform
deviate  yielding less symmetric, more edge-leading profiles than
those models with a Gaussian deviate.   Owing to the uniform deviate,
the strongest peak is distributed uniformly within the velocity 
interval.  As a result the model
yields acceptable values of $P_{KS}(f_{edg})$ for a wide
range of cloud numbers.  However, the profiles are
still too symmetric for consistency with the Mean-Median Test
for all but the smallest values of $n_{c}$. And the model fails 
the Two-Peak Test and, the Velocity Interval Test for all values of $n_{c}$. 
Therefore, this model is highly unlikely to represent the data.
As in our treatment of the dwarf galaxies, we also considered
a model (not plotted in Figure~\ref{fail}) in which $\delv$ is 
drawn uniformly from
a distribution designed to give an acceptable distribution of
$\Delta v$.
Although this model yields $P_{KS}(\delv) \approx 1$, it is ruled out at
high confidence levels by 
both the Mean$-$Median and Two$-$Peak Tests.

\section{PROPERTIES OF ROTATING DISKS RESPONSIBLE FOR 
THE DAMPED {\lya} SYSTEMS}

Apart from rapidly rotating ``cold'' disks and slowly rotating
``hot'' disks, every 
model we tested is ruled out at higher than 99.9 $\%$
confidence by one or more of the tests.
Because the ``hot disks'', i.e., the turbulent dwarf galaxies, 
are ruled out
at $97\%$ confidence, the only considered models 
compatible with the kinematic data  are those 
with rapidly rotating ``cold'' disks.  
In order to place more rigorous bounds on parameters such as
rotation speed and disk thickness, we 
investigate the parameter space of these models. 
In particular we perform
Monte Carlo runs for a wide range of the parameters
$v_{rot}$, $h/R_{d}$, and $N_{\perp}(0)$.
Although we compute all four test statistics for each run,
only the $\Delta v$ distributions are particularly sensitive to
variations of  $ v_{rot}$, $h/R_d$,  and $N_{\perp}(0)$.
Therefore, we confine the discussion to the $\Delta v$ distributions,
and to models with
identical disks.

\begin{figure}
\centering
\includegraphics[height=4.0in, width=3.0in]{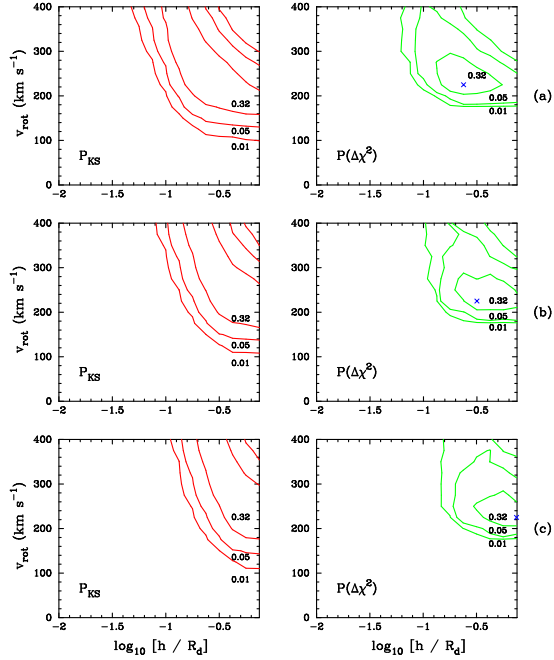}
\caption{Left Panels show isoprobability contours from the $P_{KS}({\Delta v})$
tests in the $v_{rot}$, $h/R_d$ plane for the cold rotating disk model with
$\Nperp = $ (a) $10^{20.8} \cm2$, (b) $10^{21.2} \cm2$, and 
(c) $10^{21.6} \cm2$.
Contour levels are drawn at $P_{KS}({\Delta v}) $ = 0.01, 0.05, and 0.32.
Right panels show isoprobability contours derived from the Likelihood
Ratio Test with the same contour levels.  Note that disks with rotation
speeds $v_{rot} < 180$ \kms or thickness $\log [h/R_d] < -1.0$
are extremely unlikely.}
\label{rsltfig}
\end{figure}

The results are shown in 
Figure~\ref{rsltfig} (left side) as iso-probability
contours in the $h/R_d$, $v_{rot}$ plane
for $N_{\perp}(0)$  = (a) $10^{20.8} \cm2$, (b) $10^{21.2} \cm2$, and
(c) $10^{21.6} \cm2$.  The contours correspond to selected values of
$P_{KS} (\delv)$, the KS probability discussed in $\S~\ref{rstssec}$.
Because the KS test is better suited for ruling out rather than 
establishing null hypotheses, these contours indicate the range of 
$v_{rot}$ and $h/R_d$ excluded by the data; in particular, thin
disks with slow rotation are unlikely. 

\break

To find the most likely range of parameters, 
we turn to the Likelihood Ratio Test.  Define 
the likelihood,

\begin{equation}
L \equiv \prod\limits_{i=1}^{n_{d}} p (\delv_{i}) \cmma
\end{equation}

\noindent where $p (\delv_{i})$ is the probability a model yields
the velocity interval, $\delv_{i}$, assigned to the $i^{th}$ member
of a sample comprising
$n_{d}$ damped \lya systems.
We compute $L_{max}$, the maximum value of $L$ with respect to the
parameters, i.e., we maximize the function $L = L(v_{rot}, h/R_d,
\Nperp)$ in the $v_{rot}$, $h/R_d$ plane or the $v_{rot}$,
$\Nperp$ plane.  For the $v_{rot}$, $h/R_d$ plane the maximal points 
are indicated by the $\times$'s on the
right side of Figure~\ref{rsltfig}.
We then compute the probability $- 2 \ln (L / L_{max})$
which is asymptotically distributed as $\Delta \chi^2$ for 2 degrees of
freedom in either plane, where $\Delta \chi^2$ = $\chi^2$ $-$
$\chi^2_{min},$ and $\chi^2_{min}$ is $\chi^2$ at the maximal point.
The contour levels
on the right side of Figure~\ref{rsltfig} are the confidence levels
corresponding to $\Delta \chi^2 \geq - 2 \ln (L / L_{max})$ for
$N_{\perp}(0)$  = (a) $10^{20.8} \cm2$, (b) $10^{21.2} \cm2$, and
(c) $10^{21.6} \cm2$. When $\Nperp = 10^{21.6} \cm2$, 
$L_{max}$ lies on the edge of the parameter space we investigated
and as a result determination of the ``true''
maximum is uncertain. We do not consider configurations
with $\log [h/R_d] > -0.25$  as they do not qualify  as disks.
Therefore, it is possible that the contours with this $\Nperp$
will encompass an even smaller area in this parameter space.

\begin{figure}
\centering
\includegraphics[height=4.0in, width=3.0in,angle=-90]{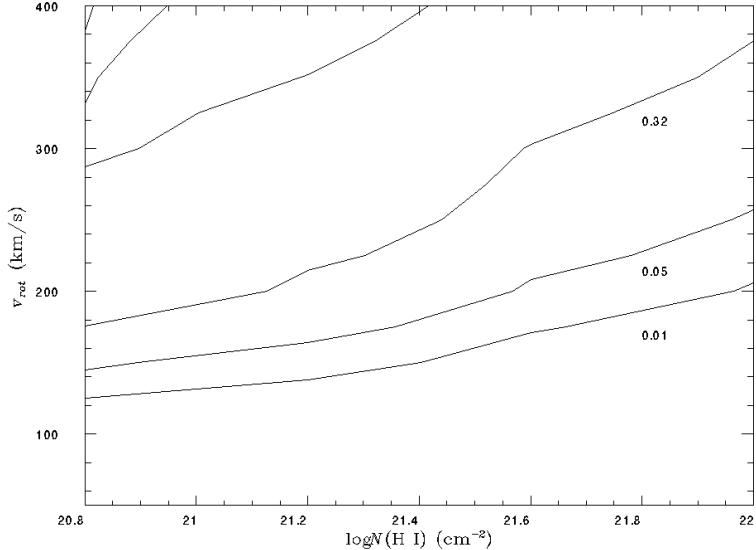}
\caption{Isoprobability contours of $P_{KS}({\Delta v})$ values in 
the $v_{rot}$,
$\N{HI}$ plane with $h/R_d = 0.3$.  The contour levels are drawn at
$P_{KS} =$ 0.01, 0.05, and 0.32.  The figure demonstrates 
that higher rotation speeds are required as $\Nperp$ increases.}
\label{vNHIh3}
\end{figure}

The Likelihood Ratio Test results in several significant trends. 
First,  a lower limit of $v_{rot} > 180$ \kms
is established at the $99\%$ confidence level, 
independent of $N_{\perp}(0)$ and $h/R_{d}$. 
Upper limits on $v_{rot}$ 
are less restrictive, especially for smaller values of
$h/R_d$ and larger $\Nperp$.  In fact there is a portion of 
parameter space  allowing $v_{rot} > 375$ \kms at the
95$\%$ confidence level for all 3 values of $\Nperp$.
But in every  case the
optimal rotation speed is given by $v_{rot} \approx$ 225 {\kms}.
Also note that the $95\%$ and 99$\%$ contours have larger areas
when $\Nperp = 10^{20.8} \cm2$, a trend
caused by the  ``column-density kinematic'' effect discussed in 
$\S~\ref{MCmthds}$. This is further illustrated
in Figure~\ref{vNHIh3} which is an iso-probability contour plot of
$v_{rot}$ vs. $\Nperp$ for $h/R_{d}$ = 0.3.
The gradual rise of $v_{rot}$ with increasing $\Nperp$
indicates larger rotation speeds are required to compensate
for the narrower profiles produced when $\Nperp$ is large.
Second, the optimal $h/R_d$ value is at
0.32.  While we find a lower limit of 
$h/R_{d}$ $>$ 0.1 at the 99$\%$ confidence level,
the test places no upper limit on
$h/R_{d}$. This is because the gradient of $v_{rot}$
along the line of sight increases with
$h/R_{d}$ until it remains nearly constant at $h/R_{d}$ $\ge$ 0.4. 

Finally, we consider the robustness of our results, in particular
the optimal rotation speed, $v_{rot} = 225$ \kms,  found from
maximum likelihood techniques.  First, we believe the rotation
speed must be large simply because there is no evidence that
the $\delv$ with the largest values are spurious outliers.  Rather,
the distribution out to large $\delv$ contains a tail which is real;
for example, 6 of the 17 profiles have $\delv > 120$ \kms.  However,
our single population model with $v_{rot}$ the same for all
disks is naive in that it cannot explain the detection of a single
system with $\delv > 225$ \kms.  In a more realistic multi-population
model, $v_{rot}$ would be distributed out to 350 \kms in order to
explain the largest rotation speeds measured for current Sa spirals
\cite{rub85}.  Such a model could then explain $\delv \leq 350$ \kms.
Higher values would then have to be interpreted
in terms of multiple disks or some other mechanism.

\begin{figure}
\centering
\includegraphics[height=4.0in, width=3.0in]{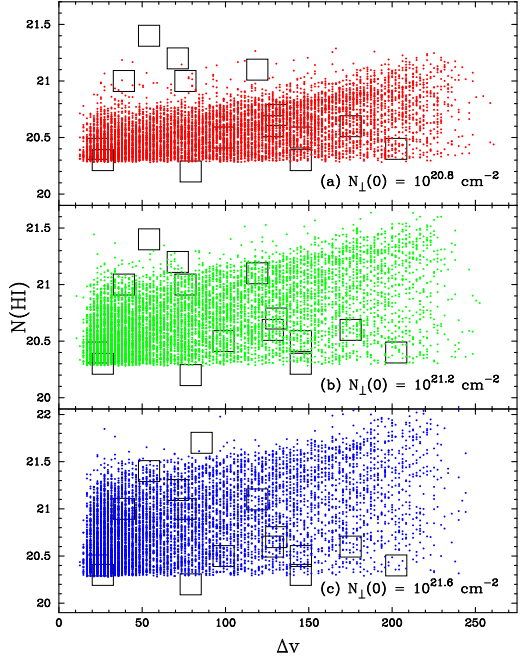}
\caption{Plot of $\N{HI}$ vs $\delv$. Clear squares represent the data. 
Dots represent simulation results for three runs with 
$h/R_d = 0.3$, and $\Nperp =$ (a) $10^{20.8} \cm2$, (b) $10^{21.2} \cm2$, and 
(c) $10^{21.6} \cm2$.}
\label{dat2D}
\end{figure}

While the statistical tests show the general features of the
``cold'' disk models are consistent with observation,
the assumption of identical $N_{\perp}(0)$ may be incorrect.
Figure~\ref{dat2D} is a plot of observed $N$(H I)
vs $\Delta v$ for 
$N_{\perp}(0)$  = (a) $10^{20.8} \cm2$, (b) $10^{21.2} \cm2$, and
(c) $10^{21.6} \cm2$.  
Although the data exhibit a possible anti-correlation between
$N$(H I) and $\Delta v$, all three models
predict a positive correlation between these quantities. 
The correlation arises because low
$\Delta v$ are generally produced at large impact parameters
where $N$(H I) is small and large $\Delta v$ are produced
at small impact parameters where $N$(H I) is large. Better
agreement with the data  could be achieved by dropping
the unrealistic assumption of identical $\Nperp$.
Thus some contribution from disks with $\Nperp$
= 10$^{21.6}$ cm$^{-2}$ would help to explain data points with
$N$(H I) $>$ 10$^{21.2}$ cm$^{-2}$, while disks with
$\Nperp$ = 10$^{20.8}$ cm$^{-2}$ help to explain the distribution
of data points at $N$(H I) $<$ 10$^{20.8}$ cm$^{-2}$. But even
this scheme produces too many disks with $\Delta v$ $<$ 60 {\kms}
and $N$(H I) $<$ 10$^{20.8}$ cm$^{-2}$. These might be eliminated 
by invoking photoionization of the outer disks where the gas density
is low and ionization parameter high.
We are currently experimenting with these and other
schemes in order to obtain a better  match with the data.

\section{AN INDEPENDENT TEST FOR RAPIDLY ROTATING DISKS}

\begin{figure}
\centering
\includegraphics[height=4.0in, width=3.0in]{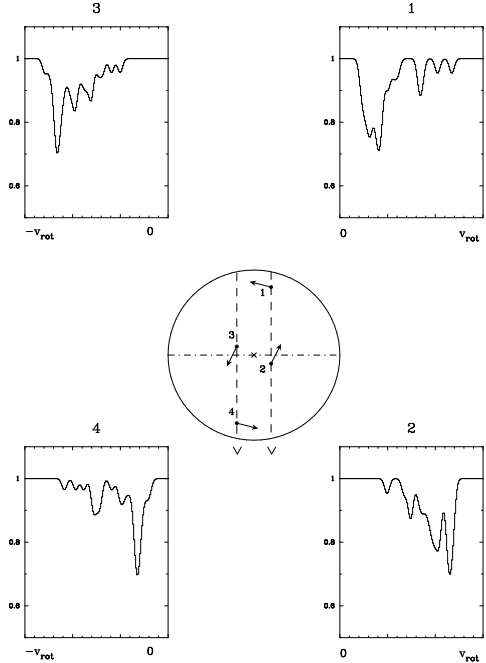}
\caption{Same as
Figure 3 except here we show 2 sightlines (dashed vertical lines).
In the profiles, the emission redshift, $v=0$, lies at
or beyond the boundaries of the absorption profile in every case
and the strongest feature lies at one of the edges.}
\label{HSTfig}
\end{figure}

Recently {\lya} emission, at $z$ = 3.15, was detected from an object
displaced by 2.3 arcsec from a sightline intersecting
a damped {\lya} system with a similar redshift 
(Djorgovski {\etal} 1996). This suggests a test of the
rotating disk hypothesis.
Let the emission
come from the center of the gaseous disk
causing absorption. 
The geometry of such a configuration is illustrated in 
Figure~\ref{HSTfig}. The circle represents a disk rotating with
$v_{rot}$ in the counterclockwise direction as indicated
by the velocity vectors. Two possible sightline geometries are indicated
by the vertical dashed lines that run from the observer
at the bottom to the QSO at the top of the figure. The kinematic major axis 
of the disk is the dash-dot-dash line. The dots at positions
1$-$4 show four locations where the sightlines encounter the disk
midplane.  The metal-line velocity profiles in the figure were
computed according to the algorithms of $\S$ 3.2 which
describes why ``edge-leading'' 
profiles arising from midplane encounters far from 
the major axis, as in positions 1 and 4, 
exhibit asymmetries that mirror ``edge-leading'' profiles arising
from midplane encounters near the major axis, as in positions 2 and 3.
The figure shows the emission redshift,
$v$ = 0, lies at or beyond the boundaries of the absorption profile
in every case. Furthermore, the component with peak optical
depth  should be at the profile boundary nearest to
$v$ = 0, when the point of midplane encounter
is far from the major axis or farthest from $v$ = 0, when the midplane 
encounter is nearer to the major axis.

As a result, the disk model predicts a systematic relation between 
the kinematics of emission and absorption lines.
By contrast the peak component 
would be {\em randomly} distributed
with respect to $v$ = 0, 
were damped {\lya} absorption to arise in undetected satellite
galaxies bound to massive primary galaxies detected in emission.
Suppose the speed of the satellite with respect to the primary
is given by $v_{sat}$, and the internal kinematics of the satellite generates
a profile with velocity interval, $\delv$. The probability
that the emission redshift of the primary is within $\delv$
is then given by

\begin{equation}
P_{profile} = (1/2)\left\{
\begin{array}{ll}
{\Delta v}/v_{sat}, & {\Delta v} < v_{sat}\\
1, & {\Delta v} \ge v_{sat}.
\end{array}  
\right.  
\end{equation}

\noindent Therefore,
when $\delv$ $<<$ $v_{sat}$, the satellite-primary configurations
are unlikely to produce emission redshifts within the absorption profile,
and are thus indistinguishable from absorption by a single large disk.
But when $\delv$ $\approx$ $v_{sat}$,
the emission redshifts are likely to lie within $\delv$. 
Because $v_{sat}$ $>$ 100 {\kms} \cite{zar95},
a significant fraction of
absorption profiles with $\delv$ $>$ 100 {\kms} 
should contain the emission redshift,
if the satellite/primary galaxy hypothesis is correct.
Clearly, the combination of  emission and absorption kinematics provides
powerful and independent tests of the rotating disk model.

Such a test
was initiated by 
Lu {\etal} (1996b). They used HIRES to obtain low-ion profiles  
for the $z_{abs}$ = 3.15 damped {\lya} absorber
toward Q2233$+$131, the same system Djorgovski {\etal} (1996)
found to be associated with an emitting
galaxy at the absorption redshift.
Comparison between emission and absorption profiles 
shows the centroid of the emission line is located outside the
absorption profile  characterized by $\delv$ = 200 {\kms}. 
\noindent This is consistent with the kinematics of a large, rotating
disk centered on the {\lya} emitter. Because the peak 
component is at the profile boundary farthest from $v$ = 0, 
the point of midplane encounter should be near
the major axis.  Lu {etal} (1996b)  used their measurement of $\delv$ 
to set a lower limit of 200 {\kms} on
$v_{rot}$ which was combined 
with the impact parameter, $b$ = 30$h_{50}^{-1}$ kpc, 
to deduce a total mass for the galaxy
of $M_{gal}$ $>$ $3 \times 10^{11} h_{50}^{-1} \rm M_{\odot}$
(note this mass is somewhat higher than the \cite{lu96b} estimate,
since we assume $\Omega_{0}$ = 0.1 rather than $\Omega_{0}$ = 1.0). 
On the other hand the
combination of large values for $b$ and $\delv$ raises potential problems
for the big disk interpretation, because the column-density velocity effect
($\S$ 3.2) indicates small $\delv$ when $b/R_{d}$ is large. Perhaps $b/R_{d}$
is small 
in this case because $R_{d}$  exceeds 10 kpc, as expected from 
the high incidence of damped {\lya} systems along
the line of sight. Moreover a large inclination angle,
$i$ $>$ 80$^{0}$, would also increase $\delv$.

\section{CONCLUDING REMARKS}

Our principal result is the CDM scenario for the formation
of galactic disks predicts rotation speeds at $z$ $\approx$ 2.5
that are ruled out at high confidence levels by the 
neutral-gas kinematics
in damped {\lya} systems. Specifically,
the semi-analytic model of Kauffmann (1996) 
predicts most precursors of current massive spirals to be 
disks embedded in dark-matter halos with masses less than 
$\approx$ 10$^{10}$$M_{\odot}$
and rotation speeds less than 100 {\kms}. By
contrast, the likelihood ratio test in $\S$ 8 requires minimum
rotation speeds, $v_{rot}$ $>$ 180 {\kms}, at the 99 $\%$ confidence 
level. The most likely value, $v_{rot} = 225 \rm \; km \; s^{-1}.$
While estimates of the masses responsible
for rotation are not generally available, 
the $3 \times 10^{11} h_{50}^{-1} \rm M_{\odot}$ 
lower limit set for the galaxy at
$z$ = 3.15 is an important constraint, because galaxies with
such large masses rarely contain disks with significant
gaseous cross sections  at large redshifts in hierarchical
scenarios \cite{wht91,kau96}.
Rather our impression is that disks embedded in 
low-mass halos causing low rotation speeds dominate the damped {\lya} interception
probabilities in most hierarchical  models.
We will check this impression by using our Monte Carlo techniques to test
the predictions of other independent CDM models (see the numerical
models of \cite{ktz96}) and Hot $+$ CDM models \cite{kly95}.
Furthermore, we will test the recent suggestion that damped
{\lya} absorption arises in multiple low-mass galaxies rather than
in a single object \cite{rau96,grd96}.
However, until such schemes are shown to
be consistent with the kinematic data, we shall adopt
the idea of large, rapidly rotating disks that formed prior to  $z$ $\sim$ 3
as a working hypothesis for the damped {\lya} systems.
We note 
the early formation of  large disks in massive galaxies is
a signature of the isocurvature models \cite{pbl93,pbl96},
and possibly hierarchical models where $\Omega_{0} \ll 1$
and $\Lambda = 0$.

Let us take a closer look at the rapidly rotating disk model.	
First, we assumed the rotation curve was flat
for all radii. This is true in current spirals,
owing to a coincidence at small radii of 
(i) the sum of the radial gravitational forces due to 
stellar disk and bulge,
and (ii) the radial gravitational force due to the dark-matter halo
\cite{bin87}.
However, in damped {\lya} systems the force due to
the disk is small compared to the halo
or bulge
(the evidence for bulges in damped \lya systems stems from the
recent detection of damped \lya absorption, at the emission redshift,
in the spectrum of a Lyman-break  bulge at $z$ = 2.7 \cite{ste96b}),
because 
the surface densities of the gaseous damped {\lya} disks
are low compared
to the surface densities of stellar disks in current spirals.
For example
the central surface densities of disk-population stars in spirals like the Galaxy 
are typically 500 to 1000 $M_{\odot}$pc$^{-2}$, whereas
the central surface densities of gas in damped {\lya}
systems are at least 10 times smaller (see \cite{wol95a}).
As a result rotation
curves in damped systems should behave as follows.
In the absence of a bulge, the rotation speed due to a spherical
isothermal halo rises out to the core radius and remains flat at larger radii. 
The presence of a bulge in the damped \lya system
results in a steeper
rise from $R$ = 0 because of the smaller core radius of the bulge. At
larger radii
the rotation speed will drop and then rise again at radii sufficiently
large for the halo to dominate. Sightlines penetrating these types of
disks may produce
$\delv$ smaller than predicted by our model. 
We are currently constructing self-consistent models 
including bulge, disk, and halo components. Second,
we assumed $v_{rot}$ was independent of $Z$.
In a self-consistent model the azimuthal velocity, $v_{\phi}(R,Z)$,
would decrease with increasing $Z$ because the $R^{th}$ component
of the radial force decreases with $Z$ for spherically symmetric
mass distributions. We carried out some preliminary tests of this
effect by imposing a negative $Z$ gradient on $v_{rot}$.
Specifically, we let $v_{rot}$ vary linearly from 225 {\kms} at
$Z$ = 0 to 150 {\kms} at 1 kpc and found the gradient had only minor
effects on the test statistics. Third,
we are investigating the limits ionizing radiation places on the sizes of the
gaseous disks and its effect on the test statistics.

Having explored the gas we turn to the stars by speculating about
possible stellar byproducts of the damped {\lya}
disks.  The kinematic evidence points to 
disks that: (1) are rapidly rotating, $v_{rot}$ $\approx$
225 {\kms}; (2) are ``cold'', $\sigma_{cc}$ $<<$ $v_{rot}$;
and (iii) may have
large scale heights, since $h$ $>$ 0.1$R_{rad}$.
By contrast the abundance measurements bring to
mind ``hot'' configurations in which $\sigma_{cc}$
$>>$ $v_{rot}$. This is because the low metallicity, 
[Zn/H] = $-$1.2 \cite{ptt94} or [Fe/H] = $-$ 1.5 \cite{lu96b}, and
the  relative abundance patterns, suggesting gas enrichment
only by type II supernovae \cite{lu96b}, 
are signatures of halo population II stars. 
Therefore the kinematic and chemical evidence are in
conflict.

The conflict might be resolved if star formation in damped
{\lya} disks results in 
a thick, metal-poor disk not unlike
the thick stellar disk of the Galaxy \cite{gil89}. The latter
is characterized by scale height $h$ $\approx$ 1.5 kpc.
The scale height is compatible with $h$ $>$ 0.1$R_{d}$ inferred for
the damped systems because $R_{d}$ for current
spirals ranges between 3 and 6 kpc \cite{kent86}, and  $R_{d}$ for
high-$z$ damped systems
must be more than twice these values to explain 
their incidence along the line of sight \cite{lzt93}.
However, it is not obvious that the assumed velocity dispersion,
$\sigma_{cc}$ = 10 {\kms},
\noindent can maintain matter, gas or stars, at scale heights
as large as 1.5 kpc.
At the solar neighborhood, where the surface density
$\Sigma$ = 70 $M_{\odot}$pc$^{-2}$, stars with
$\sigma_{cc}$ $\approx$ 25 {\kms} have a scale height $h$ = 0.4 kpc
\cite{mihalas81}.
Because the
disk models fail the Velocity Interval and Two Peak tests
when $\sigma_{cc}$ $>$ 25 {\kms}, large physical scale heights
could be a serious problem for the disk model.
However, the average surface density for damped {\lya} systems
is only
$\approx$ 10$M_{\odot}$pc$^{-2}$, and since 
$\sigma_{cc}$ $\propto$ $(h{\Sigma})^{1/2}$ for self-gravitating
disks,
$h$ $\sim$ 1.5 kpc is in principal compatible with $\sigma_{cc}$ $<$
20 {\kms}. On the other hand, the damped {\lya} disks may 
be self-gravitating only at $R$ $<$ $R_{d}$. This is because
the vertical structure of the disk is controlled
by dark matter in the halo rather than self gravity
at $R$ $>$ $R_{d}$ for exponential disks with
central surface densities low compared to that of the Galaxy (Gunn 1982).
In the dark-matter limit
$h$ $\approx$ $\sigma_{cc}$$R$/$v_{rot}$
and $\sigma_{cc}$ $\approx$ 10 {\kms}. 
As a result thick disks with low velocity
dispersions may be physically plausible at any radius.
While we have
not run simulations in which $h$ $\propto$ $R$, we see no
reason why ``flaring'' disks would dilute the
asymmetries required to explain the line profiles.

Other parameters characterizing the thick stellar disk are
the mass fraction, rotation speed, and metallicity. The 
mass fraction,
$M_{thick \ disk}/M_{thin \ disk}$ $\approx$ 0.1.
This  could be explained by a 10 $\%$
decrease of $\Omega_{g}(z)$ between $z$ = 3 and 1.5, which is consistent
with the observed variation of $\Omega_{g}(z)$ \cite{stor97}.
The rotation speed is given
by $v_{rot}$ $\approx$ 150 {\kms}.
Although this is
lower than the optimal rotation speed,
$v_{rot}$ = 225 {\kms},
a vertical velocity gradient going from
$v_{rot}$ = 225 {\kms} at midplane to 150 {\kms} at $Z$ $\approx$ $h$
would explain the kinematics and results in simulated
test-statistics compatible with the data, as discussed above. 
The most difficult problem
has to do with  metallicity, because the average [Fe/H] for the
damped systems is between a factor of 4 and 10 below the mean metallicity
of the thick disk, $\langle$[Fe/H]$\rangle$ = $-$0.6 \cite{gil89}. 
However, the thick disk abundance data refer only to stars near the solar
circle, whereas the relevant metallicity 
is an average over the entire
thick disk, which is unknown. As a result 
[Fe/H] averaged over thick stellar
disks would range between $-$1.2 and $-$1.6 in this scenario. 

Finally we cannot rule out the possibility that
a significant fraction of the neutral gas experiences negligible
star formation between $z$ = 3 and 1.5. In that case the descendents
of damped {\lya} systems would be extended metal-poor configurations
of gas surrounding the visible parts of galaxies. The detection
of low metallicity damped {\lya} systems at low redshifts, $z$ $<$ 1,
provides evidence
for such gas \cite{mey95}.  In this case the agreement between  
the comoving density of neutral gas at $z$ $\approx$ 3 and
the density of visible matter in current galaxies
would refer to the inner disks where star formation occurs, presumably
at $z$ $<$ 1.5.

\vskip 0.5truein

We are deeply indebted to
the group headed by W. L. W. Sargent for generously providing us with
their HIRES spectra. 
We are especially grateful to Limin Lu
for extensive discussions and comments.
The authors would also
like to thank Tom Barlow for his excellent HIRES data 
reduction software.
We also thank Kim Griest, Ken Lanzetta, 
Jim Peebles, Joe Silk, David Tytler, and Amos 
Yahil for stimulating discussions.  
AMW and JXP were partially supported by 
NASA grant NAGW-2119 and NSF grant AST 86-9420443.

\clearpage

\end{document}